\documentclass[10pt]{article}
\usepackage{jheparxiv}
\usepackage{bbold}
\usepackage{amsmath}
\usepackage{amsthm}
\usepackage{amssymb}
\usepackage{amsfonts}
\usepackage{tensor}
\usepackage{braket}
\usepackage{slashed}
\usepackage{graphicx}
\usepackage{algpseudocode} 
\usepackage{tikz} 
\graphicspath{dmathesis-masterv2/}

\theoremstyle{definition}

\newcommand{\cxh}{\mathbb{C}\times\mathbb{H}}
\newcommand{\vect}{\textbf}
\newcommand{\Tr}{\text{Tr}}
\newcommand{\Trr}{\text{Tr}_{2}}
\newtheorem{theorem}{Theorem}[section]

\newtheorem{lemma}[theorem]{Lemma}

\newcommand{\imh}{\text{Im}_{\mathbb{H}}}
\newcommand{\imc}{\text{Im}_{\mathbb{C}}}
\newcommand{\reh}{\text{Re}_{\mathbb{H}}}
\newcommand{\rec}{\text{Re}_{\mathbb{C}}}
\newcommand{\cO}{\mathcal{O}}
\newcommand{\cL}{\mathcal{L}}

\newcommand{\var}[1]{\textit{#1}}

\title{Notes on the Dynamics of Noncommutative U(2) and Commutative SU(3) Instantons}
\author{Douglas J. Smith,}
\author{Calum J. Robson,}
\author{Joseph A. Farrow}

\affiliation{Department of Mathematical Sciences, Durham University, Durham, DH1 3LE, United Kingdom}
        
\emailAdd{joseph.a.farrow@durham.ac.uk}
\emailAdd{c.j.robson@lse.ac.uk}
\emailAdd{douglas.smith@durham.ac.uk}

\abstract{
We examine the dynamics of noncommutative instantons of instanton number $2$ and commutative instantons of instanton number $3$ in 5d Super Yang Mills theory. We begin by detailing the construction of the 1/4-BPS instanton solutions, their moduli space, and the moduli space potential using an explicit parametrisation of the moduli space coordinates in terms of the biquaternions. We then go on to numerically analyse the dynamics on the moduli spaces we have constructed, discussing some of the numerical issues which arose and describing the code we developed to solve them}

\begin{document}

\maketitle

\section{Introduction}

The aim of this paper is to present new solutions for the moduli space dynamics of 2- and 3- instantons in 5d SYM theory.  Instantons are a specific example of topological solitons, which are nonlinear solutions to certain PDEs. Because the properties of these solutions are tied to topological invariants of the spaces they are defined upon, they are very stable -- no continuous transformation (including time evolution) of the solutions can cause these properties to change. Originally discovered in \cite{Belavin1975}, research into their properties took off after the discovery of the ADHM method for constructing them in \cite{Atiyah1978}. Whilst instantons were originally constructed as solutions in 4 Euclidean dimensions, we can also define them in 5d SYM. Here the instantons are static solutions in every slice of the 4 Euclidean dimensions, and the 5th timelike dimension is seen as describing their evolution. In the context of string theory this arises since the instantons appear as 1/2-BPS states corresponding to an interacting system of D0-branes and D4-branes \cite{Witten1996} \cite{Tong2005}.\\
In this system, there are have 5 $SU(N)$ scalar fields which describe the transverse positions of the D4-branes. Separating the branes gives at least one of these scalars a non-zero expectation value. In the low energy limit this corresponds to introducing a non-zero scalar field on top of the instanton equations. This configuration would usually be an unstable field configuration due to interactions with the Higgs field however  introduction of the scalar field gives the instantons an electric charge which balances the scalar Higgs field, producting a stable solution.\cite{Lambert1999}\cite{Peeters2001}. These charged instantons are known as dyonic instantons, and are 1/4-BPS rather than 1/2-BPS. They are the low energy limits of a bound state of fundamental strings and D0-branes.  \\
This system has been the subject of particular interest as the low energy limit of M5 branes in M-Theory. The low energy dynamics of these objects is described using the so-called (2,0) theory, a 6d Superconformal field theory. It has been shown that if we dimensionally reduce this theory, we get 5d SYM \cite{Aharony1998}\cite{Aharony19982}. Another way of reducing the number of dimensions is to compactify one of the 6 dimensions  into an $S^{1}$. Somewhat surprisingly, it turns out that this also gives 5d SYM theory. In this case, the instanton sector, with separate solutions labelled by the integer instanton number $k$,  has been shown to agree for low $k$ with the Kaluza-Klein modes arising from the compactifaction, which are labelled by an integer winding number. This raises the possibility that if these sectors are in fact identical, then 5d SYM with all instantons included corresponds to including all the Kaluza-Klein modes of the compactified (2,0) theory. This would imply that the 6d (2,0) theory is the UV fixed point of 5d SYM, even though 5d SYM is pertubatively non-renormalisable\cite{Douglas2011}\cite{Lambert2011}. \\
Directly calculating the dynamic behaviour of instanton solutions is both analytically and computationally expensive. The moduli space method developed by Manton \cite{Manton1981} simplifies things by treating the free parameters as coordinates on a manifold, called the moduli space. Evolution of an instanton solution is approximated for slow motion by geodesic motion on the moduli space. This moduli space contains singularities when the size of the instanton shrinks to zero size. They correspond in the string theory to a transition between the Coloumb and Higgs branches of the D0 theory \cite{Tong2005}. \\
This is the motivation for introducing noncommutativity, corresponding to adding a Fayet-Illiopoulos term to the string theory. Defining the instantons on a noncommutative spacetime has the effect of introducing a minimum size for the instantons. This resolves the moduli space singularities as the instantons can only shrink to a finite size, so cannot reach the singularity whee their size vanishes.\\
We begin by introducing instantons themselves, and the noncommutative spacetimes we will be studying them on.  This includes a brief discussion of the biquaternions -- the algebra $\mathbb{C}\times\mathbb{H}$. The calculation of instanton solutions uses the ADHM construction first developed in \cite{Atiyah1978}. We briefly review the use of this method in the $SU(2)$ Yang Mills case.  \\
We then introduce the instanton moduli space \cite{Manton1981}. We recap how this construction can be extended to dyonic instantons via introducing a potential on the Moduli space, following the presentation in \cite{Dorey2002}. A practical method for calculating the moduli space metric and potential for noncommutative $U(N)$ instantons is presented in Appendix \ref{MethodMetric} .  This generalises that presented for $SU(2)$ commutative instantons in \cite{Allen2012}. \\

In the second part of the paper, we look for solutions to the equations we have derived. The noncommutative two-instanton case was first studied in \cite{Iskauskas2015} however we found an error in this result. We were unable to find the exact result for the full moduli space but we present a solution defined on a subspace of the full moduli space taking values in the $\mathbb{C}\times\mathbb{C}$ subgroup of $\mathbb{C}\times\mathbb{H}$. This is a geodesic submanifold of the full moduli space. After finding this solution, we use it to derive the metric and potential on this subspace. We then to numerically evaluate scattering in this subspace. We compare these results to the results for the commutative two-instanton in \cite{Allen2012}. \\
Finally, we look at the commutative three-instanton case. Again we present a solution for the complex subspace $\mathbb{C}\times\mathbb{C}$ and calculate the moduli space metric and potential for that subspace. Numerical scattering calculations proved to be very computationally expensive, however we were able to plot scalar field and topological charge density profiles, which allowed us to make some comparison to the two instanton case in the appropriate limits. \\

We use two different numerical algorithms to integrate the moduli space equation of motion, depending on the algebraic complexity of the moduli space metrics and potentials considered. For two non-commutative instantons with an orthogonal gauge embedding and 4 dimensional moduli space, we find the system sufficiently simple to solve with the numerical algorithm developed in~\cite{Allen2012}. Increasing to the full 6 dimensional gauge embedding results in significantly more complex metrics and potentials on the moduli space, and we find that these cases are no longer tractable to the algorithm from~\cite{Allen2012}. To overcome this problem we develop a new numerical algorithm, which we describe in Appendix~\ref{App:numerics}. We will attach a {\sc Mathematica} notebook containing the methods discussed in this appendix to a future arXiv submission of the paper, which can be accessed by following the links `Other formats' and then `Download Source'. 

\section{Background Material}
We begin by defining our notations and conventions for the biquaternions and for noncommutative spacetime. These will be used throughout the rest of the paper. 
\subsection{Quaternions and Biquaternions}\label{sec:biquaternions}
The group $\cxh$, known as Complex Quaternions, Biquaternions and even Tessarions has a long history \cite{Hitzer2012}. To avoid confusion we will refer to the group as Biquaternions in the rest of the paper.  As discussed in, e.g. \cite{Francis2005}, the algebra is equipped with three notions of conjugation. We write a general element of the group as
\begin{equation}
q=q_{R}+iq_{I}=q_{R0}+\vect{q}_{R}+i(q_{I0}+i\vect{q}_{I})
\end{equation}
Where $q_{R}, q_{I}\in\mathbb{H}$, and correspondingly $q_{R0}, q_{I0}\in\mathbb{R}$ and $\vect{q}_{R}, \vect{q}_{I}$ belong to the quaternion imaginary part of $\mathbb{H}$. Then we have a complex conjugation $q^{\star}$, which takes
\begin{equation}
q_{R}+iq_{I}\rightarrow q_{R}-iq_{I}
\end{equation}
We also have a quaternion conjugation $\bar{q}$
\begin{equation}
q_{R}+iq_{I}\rightarrow\bar{q}_{R}+i\bar{q}_{I}=q_{R0}-\vect{q}_{R}+i(q_{i0}-\vect{q}_{I})
\end{equation}
Finally we have a total conjugation $q^{\dagger}$ which applies both these operations simultaneously
\begin{equation}
q_{R}+iq_{I}\rightarrow\bar{q}_{R}-i\bar{q}_{I}=q_{R0}-\vect{q}_{R}-i(q_{i0}-\vect{q}_{I})
\end{equation}
To clarify our notation,  we use the basis $\sigma_{n}$ for the quaternions, with $\sigma_{n}=(\mathbb{1}_{2}, i\tau_{i})$, where the $\tau_{i}$ are the standard Pauli matrices. We also define $\bar{\sigma}_{n}=(\mathbb{1}_{2}, -i\tau_{i})$. 
Further, we define the selfdual object
\begin{equation}
\sigma_{mn}=\frac{1}{4}\Big(\sigma_{m}\bar{\sigma}_{n}-\sigma_{n}\bar{\sigma}_{m}\Big)
\end{equation}
and the anti- selfdual
\begin{equation}\label{doublesigma}
\bar{\sigma}_{mn}=\frac{1}{4}\Big(\bar{\sigma}_{m}\sigma_{n}-\bar{\sigma}_{n}\sigma_{m}\Big)
\end{equation}
Here, we define a selfdual matrix as one for which $A_{nm}^{\star}=A_{mn}$, and an anti-selfdual one as $A_{nm}^{\star}=-A_{mn}$. With these definitions, we have
\begin{equation}\label{qbasis}
\sigma_{0}=	\begin{bmatrix}
1 & 0\\ 0 &1
\end{bmatrix}, \ \ 
\sigma_{1}= \begin{bmatrix}
0 & -1 \\ 1 & 0
\end{bmatrix}
\ \ 
\sigma_{2}= \begin{bmatrix}
0 & i \\ i & 0
\end{bmatrix}
\ \ 
\sigma_{3}= \begin{bmatrix}
i & 0 \\ 0 & -i
\end{bmatrix}
\end{equation}
Finally, the fact that the biquaternions have multiple notions of conjugation means that there are multiple notions of the imaginary part. We will use $\imh$ to denote the quaternion imaginary part, defined for $q=q_{0}+\vect{q}\in\mathbb{H}$, with $q_{0}\in\mathbb{R}$ and $\vect{q}$ a purely imaginary quaternion, as
\begin{equation}
\imh(q)=\vect{q}
\end{equation}
We also have the complex imaginary part, defined for $z\in\mathbb{C}$, $z=x+iy$ as
\begin{equation}
\text{Im}_{\mathbb{C}}(z)=y
\end{equation}
Note that $\imc$ doesn't include the factor $i$ which we must add in by hand where it is required -- this is done to match with the usual definition of Im in the complex case, however it does mean some care has to be taken when restricting from $\mathbb{H}$ to $\mathbb{C}$, as $i$ then corresponds to the imaginary quaternion basis vector, which is included in $\imh$ but not in $\imc$.\\
In the case of a biquaternion $q=q_{R}+iq_{I}$ we have
\begin{equation}
\imc(q)=q_{I} \ ; \ \imh(q)=\vect{q}_{R}+i\vect{q}_{I}
\end{equation}
where $\vect{q}_{R}$ and $\vect{q}_{I}$ are the quaternion imaginary parts of $q_{R}$ and $q_{I}$ respectively. We similarly define $\reh$ and $\rec$.
\subsection{Noncommutativity}\label{Noncomm}
It is convenient to introduce the study of noncommutative spacetimes into the study of instantons. It is convenient because it allows us to resolve singularities on the instanton moduli space (see the next section). It was first show in \cite{Nekrasov:1998ss} that this was possible, and since then many examples have been constructed (see e.g. \cite{Bak2013}\cite{Correa2001}\cite{Chu2001} for a selection).\\
To construct a noncommutative version of $\mathbb{R}^{4}$, we simply impose an anticommutation relation on the spacetime coordinates
\begin{equation}
\big[x^{m}, x^{n}\big]=\theta^{mn}
\end{equation}
Here $m,n$ are the Euclidean Lorentz indices, and $\theta^{mn}$ is a real, antisymmetric, constant matrix. We can always rotate it into the form
\begin{equation}
\theta^{mn}=\begin{bmatrix}
0& \theta^{12}&0 &0 \\
-\theta^{12} & 0 & 0 & 0\\
0 & 0 & 0 & \theta^{34}\\
0 & 0 & -\theta^{34} & 0
\end{bmatrix}
\end{equation}
There are several interesting subcases of this matrix \cite{Chu2001}; in this paper we consider the selfdual (SD) case, where $\theta^{12}=\theta^{34}=2\zeta$. \\
The noncommutativity of the spacetime coordinates forces us to modify our notion of the multiplication of functions. Rather than the usual multiplication, we use the \textsl{Moyal Star Product}  \cite{Moyal1949}. This is defined as
\begin{equation}
f(x)\star g(x)=\exp\Big(\frac{i}{2}\theta^{ij}\partial_{i}\partial'_{j}\Big)f(x)g(x')|_{x=x'}
\end{equation}
This gives the following expansion on powers of $\theta^{ij}$
\begin{equation}
f(x)\star g(x)=f(x)g(x)+\frac{i}{2}\theta^{ij}\partial_{i}f(x)\partial_{j}g(x)+\mathcal{O}(\theta^{2})
\end{equation}
Using this, the gauge potential and field strength become
\begin{equation}
A_{i}\rightarrow g^{-1}\star A_{i}\star g+g^{-1}\star \partial_{i}g
\end{equation}
and 
\begin{equation}\label{MoyalFS}
F_{ij}=\partial_{[i}A_{j]}-i\big[A_{i}, A_{j}\big]_{\star}
\end{equation}
where
\begin{equation}
\big[A_{i}, A_{j}\big]_{\star}=A_{i}\star A_{j}-A_{j}\star A_{i}
\end{equation}
This has two effects on our instanton solutions. First of all, it allows us to find solutions with no commutative equivalent, since the additional length scale $[\zeta]=[length]^{2}$ and the fact we are not in Euclidean flat space means Derrick's theorem does not apply. \\
Secondly, and less positively, in theory it implies we have an infinite number of terms to calculate. However we can avoid this thanks to an isomorphism between the algebra of functions with the $\star$-product, and certain operators over Hilbert space. This is more fully discussed in \cite{Gopakumar2000}.

\section {Dyonic Instantons}\label{DyonicADHM}
Now we have discussed these notational conventions we define what is meant by dyonic instantons. Following the presentation in \cite{Allen2012} we start with the action
\begin{equation}
S_{YM}=\int d^{5}x\frac{1}{4}F_{\mu\nu}^{a}F_{a}^{\mu\nu}+\frac{1}{2}D_{\mu}\phi D^{\mu}\phi
\end{equation}
We consider static solutions, for which the integral is taken over the four spatial dimensions. These have energy, topological charge, and electric charge given respectively by
\begin{align}\nonumber
E=&\int d^{4}x\Tr\big(\frac{1}{2}F_{i0}F_{i0}+\frac{1}{4}F_{ij}F_{ij}+\frac{1}{2}D_{0}\phi D_{0}\phi+\frac{1}{2}D_{i}\phi D_{i}\phi\big),\\ \nonumber
k=&-\frac{1}{16\pi^{2}}\int d^{4}x \epsilon_{ijkl}\Tr\big(F_{ij}F_{kl}\big),\\ 
Q_{E}=&\int d^{4}x\Tr\big(D_{i}\phi F_{i0}\big)=\int d^{4}x\Tr\big(D_{i}\phi\big)^{2}
\end{align}
We use a Bogomolny argument of the type \cite{Manton2004} to give us a bound on the energy, by completing the square
\begin{align}
E=\ &\int d^{4}x\Tr\Big(\frac{1}{8}\big(F_{ij}\pm\frac{1}{2}\epsilon_{ijkl}F_{kl}\big)^{2}+\frac{1}{2}\big(F_{i0}\pm D_{i}\phi\big)^{2}+\frac{1}{2} D_{i}\phi D_{i}\phi \\ \nonumber
& + \frac{1}{8}\epsilon_{ijkl}F_{ij}F_{kl}\mp F_{i0}D_{i}\phi\Big)
\end{align}
So we get
\begin{equation}
E\geq 2\pi^{2}\lvert k\rvert+\lvert Q_{E}\rvert
\end{equation}
where
\begin{align}
k=-\frac{1}{16\pi^{2}}\int d^{4}x\ \epsilon_{ijkl}\Tr(F_{ij}F_{kl}),\\ \nonumber
Q_{E}=\int d^{4}x \Tr(D_{i}\phi F_{i0})=\int d^{4}x \ \Tr(D_{i}\phi)^{2}
\end{align}
The conditions for this bound to be saturated are
\begin{align}\label{dyonicBPS}
F_{ij}=\frac{1}{2}\epsilon_{ijkl}F_{kl}\\ \nonumber
F_{i0}=D_{i}\phi\\ \nonumber
D_{0}\phi=0
\end{align}
With the boundary condition on $\phi$ that it goes to the vev $\phi_{0}=i\vect{q}$ at infinity \cite{Lambert1999} where $\vect{q}$ is an arbitrary imaginary quaternion. The second and third of these are satisfied provided the fields are static and $A_{0}=\phi$. Requiring our solutions to obey Gauss' law, $D_{i}E_{i}=ig\big[\phi, D_{0}\phi\big]$. imposes the further equation
 \begin{equation}\label{eq:scalarfield}
D^{2}\phi=0
\end{equation} 
So, to find a dyonic instanton solution, we can use the ADHM method to calculate a self dual field strength $F_{\mu\nu}$, then additionally calculate the background scalar field using equation (\ref{eq:scalarfield}).  An important corollary is the observation that when $\phi=0$, the solution reduces to precisely that of a pure instanton, and such solutions are in one-to-one correspondence with the pure instanton solutions discussed above. 

\subsection{The ADHM Construction}\label{ADHM}
To calculate an instanton solution, we use the ADHM construction. We follow an ansatz-based construction outlined in \cite{Chu2001}. A good discussion of this can be found in \cite{Dorey1996}.\\
The main ingredient in the ADHM construction is the ADHM Data $\Delta$. This is an $(N+2k) \times 2k$ matrix, where $N$ is the degree of the gauge group $SU(N)$ or $U(N)$, and $k$ is the instanton number, or topological degree. In the commutative case the entries are usually taken to be real, whereas in the noncommutative case they are taken as being  complex. However, they can be taken to be complex in the commutative case too, with the real solution recovered using the symmetries due to the additional redundancy. Therefore we will treat the entries as being complex in the remainder of this paper unless otherwise stated. With this in mind, we have
\begin{equation}\label{ADHMData}
\Delta=\begin{bmatrix}
\Lambda \\ \Omega
\end{bmatrix}
\end{equation}
where $\Lambda$ is an $N\times 2k$ complex matrix and $\Omega$ is a $2k\times 2k$ hermitian matrix. It is often useful for the purpose of performing calculations to treat these as being instead biquaternion-valued matrices (or quaternion-valued, for real matrices). The matrix $\Omega$ can always be treated as a $k\times k$ matrix of biquaterions, but $\Lambda$ is not as straightforward.  For some values of $N$ and $k$ there is a similar identification -- for example, for the $U(2)$ instantons we consider in this paper, we can always write $\Lambda$ as a row of $N$ (bi)quaternions. However, in general, this is not possible. \\
This difficulty is mitigated by the fact that we will always end up considering $\Delta^{\dagger}\Delta$ in any practical calculation, and as we shall see below, this can always be written in (bi)quaternion form. \\
The commutative ADHM method involves solving the equation
\begin{equation}\label{ADHMeqn}
\Delta^{\dagger}\Delta=\mathbb{1_{2}}\otimes f^{-1}
\end{equation}
where $f$ is an invertible $k\times k$ matrix, and we can think of $\mathbb{1}_{2}$ as the quaternion identity. This also means we can look at the ADHM equation above as
\begin{equation}
\text{Im}_{\mathbb{H}}\big(\Delta^{\dagger}\Delta\big)_{ij} =0
\end{equation}
where $\text{Im}_{\mathbb{H}}$ takes the quaternion imaginary part. In the noncommutative case we must modify the ADHM equation (\ref{ADHMeqn}) to be
	\begin{equation}
(\Delta^{\dagger}\Delta)_{ij}=\mathbb{1_{2}}\otimes f^{-1}_{ij}-4\zeta\sigma_{3}\delta_{ij}
	\end{equation}
which we can view as
\begin{equation}
\text{Im}_{\mathbb{H}}\big(\Delta^{\dagger}\Delta\big)_{ij} =-4\zeta\sigma_{3}\delta_{ij}
\end{equation}
This solution has some residual freedom -- we can transform any solution $\Delta$ to 
\begin{equation}
\Delta\rightarrow \ \begin{bmatrix}
1 & 0 \\
0 & R^{T}
\end{bmatrix}
\Delta \
R
\end{equation}
to obtain a new solution, where $R$ lies in $O(k)$ if we are using real quaternions, or $SU(k)$ if we are using biquaternions. This additional freedom in the biquaternion case cancels out the additional degrees of freedom from the complexified ADHM parameters. This freedom is also important in obtaining the moduli space metric (see appendix \ref{MethodMetric}). Once we have solved equation (\ref{ADHMeqn}), or its noncommutative analogue,  we can use it to calculate the gauge potential and field strength.\\
	To do this, we need to find a zero eigenvector $U(x)$ of $\Delta^{\dagger}$, normalised so that $U^{\dagger}U=1$. There are $N$ such vectors, spanning the nullspace of $\Delta$,
	satisfying
	\begin{equation}
	U^{\dagger}\Delta=\Delta^{\dagger}U=0
	\end{equation}
	This implies that $U$ has dimension $N+2k$ as a complex vector. Once we have this $U$ we can use it to define the gauge potential as
	\begin{equation}
	A_{\mu}=U^{\dagger}\partial_{\mu}U
	\end{equation}
	and the field strength as
	\begin{equation}
	F_{\mu\nu}=-4U^{\dagger}bf\sigma_{mn}b^{\dagger}U
	\end{equation}
	where $b$ is a $(N+2k)\times 2k$ matrix whose top $N\times 2k$ part is $0$ and whose bottom $2k\times 2k$ part is the identity, and $\sigma_{mn}$ is as given in (\ref{doublesigma}). This procedure works for both the commutative and the noncommutative cases. 	
	An additional subtlety in the noncommutative case comes from the assumption implicit in the above construction that we can factorise the projection operator 
	\begin{equation}
	1-U\bar{U}
	\end{equation}
	as
	\begin{equation}\label{ADHMcomp}
	1-\mathcal{P}\equiv\delta_{\lambda\kappa}\delta\indices{_\alpha^\beta}-\mathcal{P}\indices{_{\lambda\kappa}_{\alpha}^{\beta}}=\Delta f\bar{\Delta}
	\end{equation}
	This is called the, `Completeness relation'. It is automatically satisfied in the commutative case, however there are some complications in the noncommutative case. The issue is that, whereas the normalisation of $U(x)$ is straightforward in the commutative case, there is are subtleties in the case where $x$ is itself an operator. We must therefore be careful to pick a good definition for this normalisation. These issues were first discussed in \cite{Chu2001}. Checking this is highly non trivial, however as it only affects the value of $U$, and not the validity of the remainder of the solution, it is only necessary if one is constructing an explicit expression for the gauge potential, which we are not. The only point we use $U$ is in Appendix \ref{MethodPotential2}, and here we take it in the limit $x\rightarrow\infty$ in which any noncommutative effects (which go as $\frac{\zeta}{x^{r}}$ for some positive integer $r$) are automatically neglected. This does not affect our results therefore, though a full investigation might be a fruitful topic for future research. 
	\subsection{The Moduli Space}\label{subs:ModuliSpace}
	The \textsl{Moduli Space} of instanton solutions is a the space of inequivalent solutions to the self dual Yang Mills equations, (\ref{ADHMeqn}). This was first introduced for instantons in \cite{Manton1981}. Calculating the dynamics of an individual instanton solution over a period of time is difficult. However for sufficiently slow velocities we can approximate such a solution by a slow transition between different instanton solutions with marginally different initial conditions. This corresponds to motion on the Moduli Space. For a review of Techniques and applications see e.g. \cite{Dorey2002} \cite{Allen2012}\cite{Weinberg2006}\cite{Peeters2001}.
The Moduli Space is parameterised by the $4kN-4$ free ADHM parameters, which are called the Collective Coordinates. 
 To define small velocities, we must introduce a moduli space metric. To do this we look at small fluctuations $A_{m}(x)+\delta A_{m}(x)$. If this is also to be a solution to the equations (and hence lie in the moduli space), the $\delta A_{m}$ must satisfy the linearised self duality equation
\begin{equation}\label{LinYM}
\mathcal{D}_{m}\delta A_{n}-\mathcal{D}_{n}\delta A_{m}=\epsilon_{mnkl}\mathcal{D}_{k}\delta A_{l}
\end{equation}
In addition, it must not be related to $A_{n}(x)$ by a local gauge transformation. We therefore require the solutions to be orthogonal to gauge transformations. To define this orthogonality, we take the natural metric on the space of all solutions
\begin{equation}
g\big(\delta A_{m}(x), \delta A'_{m}(x)\big)= \int d^{4}x\Tr\big(\delta A_{m}(x)\delta A'_{m}(x)\big)
\end{equation}
and then use this to induce a metric on the moduli space after quotienting out the gauge-equivalent solutions. We then require that under this metric, zero modes $\delta A_{i}(x)$ are orthogonal to all gauge transformations $D_{i}\Lambda$. 
This is equivalent to satisfying
\begin{equation}
D_{i}\delta A_{i}=0
\end{equation}
 \\
 For small pertubations, we get the following action on the moduli space \cite{Nekrasov2000}\cite{Dorey2002}

\begin{equation}
S=\frac{1}{2}\int d^{5}x\Tr\big(F_{i0}F_{i0}-D_{i}\phi D_{i}\phi+D_{0}\phi D_{0}\phi\big)
\end{equation}
If we neglect terms of order $\dot{z}^{2}\lvert q\rvert^{2}$, where $z(t)$ refers to any of the collective coordinates on the moduli space, we get the effective action
\begin{equation}\label{eq:effact}
S=\frac{1}{2}\int dt \Big(g_{rs}\dot{y}^{r}\dot{y}^{s}-\lvert q\rvert^{2}g_{rs}K^{r}K^{s}\Big)
\end{equation}
Where the $K_{r}$ are Killing vectors of the moduli space and satisfy $D_{m}\phi=\lvert q\rvert K^{r}\delta_{r}A_{m}$. Then equation (\ref{eq:effact}) is the sum of a free instanton and the potential 
\begin{equation}
V=\frac{1}{2}\int d^{5}x\Tr(D_{i}\phi D_{i}\phi)=\frac{\lvert q\rvert^{2}}{2}\int dt g_{rs}K^{r}K^{s}
\end{equation}
This solution is valid in the limit
\begin{equation}
\dot{z}^{2}, \lvert q\rvert \leq 1; 
\end{equation}
where we can ignore terms of order $\dot{z}^{2}\lvert q\rvert^{2}$ and higher. The geometric interpretation of this is both that the kinetic energy of the instanton solution is sufficiently small, and that the potential evaluated on the instanton solutions, which lie on the moduli space, is shallow compared to the potential on non-BPS solutions evaluated off the moduli space. This allows us to imagine our approximate solution as lying in a steep valley given by the locally small potential around the moduli space solutions, where the small kinetic energy prevents our dynamics from, `climbing away' from the moduli space. 
\subsection{The Complex Subspace}\label{CompSub}
There is one final technical point to discuss, which applies to both Pure and Dyonic Instantons. The Moduli space has several subspaces, which are preserved under the geodesic motion. This means a geodesic beginning in one of these subspaces (i.e. whose initial tangent vector lies in that subspace) will remain in it throughout its motion.\\
The subspace we are interested in is as follows. The moduli space is a manifold over the collective coordinates $\vect{z}$. However, as we saw in section \ref{ADHM} these collective coordinates are elements of the ADHM matrix $\Delta$ and therefore are parametrised by the (bi)quaternions. That algebra can be thought of as $\mathbb{C}\times\cxh$. Therefore, in the same way that the quaternions contain an invariant complex subspace $\mathbb{C}$, the biquaternions have the subspace $\mathbb{C}\times\mathbb{C}$. We can therefore restrict from the full moduli space where the collective coordinates are biquaternions, to a submanifold where they lie in $\mathbb{C}\times\mathbb{C}$.\\
This corresponds to conjugating all the moduli space coordinates by a unit quaternion $q$, e.g. $\tau\rightarrow q\tau\bar{q}$. This corresponds geometrically to leaving the real quaternion part fixed, whilst the imaginary quaternion is rotated around an axis in $S^{3}$ represented by $q$.  Imposing invariance under such rotations corresponds to requiring our solutions to be fixed points under this rotation, which constrains them to lie in a two dimensional plane within the four dimensional quaternions. Because $\bar{q}q=1$, if we multiply two (bi)quaternions together and apply this rotation to each of them then the result is that the entire product is rotated -- e.g. 
\begin{equation}
\Delta^{\dagger}\Delta\rightarrow q\Delta^{\dagger}\bar{q}q\Delta\bar{q}=q\Delta^{\dagger}\Delta\bar{q}
\end{equation}
We can therefore think of all our equations and objects (for example the scalar field and potential) as being rotated in the same overall way. 
In the commutative space, for pure instantons we can see the invariance of this subspace automatically, since the elements of $\mathbb{C}\in\mathbb{H}$ automatically commute with all other elements, meaning that they form an ideal within that group (and ideals are invariant subspaces) \cite{Furey2016}. For dyonic instantons, we must choose the imaginary direction to be the same as the vev in SU(2), since otherwise the vev will not be preserved. The result is that the transformation maps solutions to solutions, and so the fixed point manifold thus generated is also a geodesic submanifold of our moduli space. In the noncommutative case, the presence of the noncommutative parameter means that the spacetime coordinates do not automatically commute. The only complex subspace which is preserved in this case is  the complex subspace spanned by $\{\mathbb{1}, \sigma_{3}\}$, as, since $\sigma_{3}$ is the direction associated with the noncommutativity, it is preserved under rotations of the space, and $\mathbb{1}$ commutes with everything. We must then align the plane within $\mathbb{H}$ that we are preserving with this direction. Hence we see, as would be expected, the presence of a noncommutative parameter reduces the symmetries of the theory. \\
Calculations on the full quaternion moduli space are very computationally expensive, and this subspace is often much easier to run simulations on. In addition, the fact that elements in this subspace commute makes solving the ADHM equations on this restricted part of the theory much easier.

\section{The Two Instanton Solution}\label{sec:twoinstatonsol}
Now we have discussed the technical background, we present our work on the dynamics of the noncommutative U(2) 2-instanton. First, we review the single U(2) instanton as presented in \cite{Bak2013}. This is necessary to test the two instanton solution in the appropriate limits. \\
Next we derive the ADHM equations for the two Instanton case, using biquaterion coordinates. We were unable to find a solution for the full moduli space, however we were able to find a solution for the geodesic submanifold discussed in section \ref{CompSub}. This corrects the solution in \cite{Iskauskas2015} \\
After finding this solution we use it to derive the metric and potential for the relevant moduli space, and show that the metric and potential behave suitably in the commutative limit and in the limit of the instantons being  far separated. 	
\subsection{The Single U(2) Instanton}\label{sec:the-single-u2-instanton}
First, we state the solution for a single U(2) instanton in noncommutative space. We will follow \cite{Bak2013}, however we rederive their solution in our notation. One key difference is that the relation between their $\zeta'$ and our $\zeta$ is $\zeta'=2\zeta$  Other discussions of the solution can be found in \cite{Chu2001} and \cite{Correa2001}. First, for the single noncommutative U(2) instanton, the ADHM data has the form
\begin{equation}
\begin{bmatrix}
v_{R}+v_{I}\sigma_{3} \\ X-x'
\end{bmatrix};\ v_{R}, v_{I}, X, x'\in\mathbb{H}
\end{equation}
 As discussed in \cite{Bak2013}, we can set $X-x'=0$ by an symmetry transformation related to the centre of mass. Then we can solve for $v_{I}$ in terms of the free parameter $v_{R}$. 
\begin{equation}
v_{I}=-\frac{2\zeta v_{R}}{\lvert v_{R}\rvert^{2}}; \ \ v_{R}\in \mathbb{H}
\end{equation}
This gives us 4 ADHM coordinates, as required. 
Now we have this, we can calculate the metric and potential using the method discussed in appendices \ref{ScalarField} and \ref{MethodPotential2}.
The potential is equal to 
\begin{equation}
\mathcal{V}=8\pi^{2}\lvert \vect{q}\rvert^{2}\Big(\rho^{2}-\frac{16\zeta^{2}}{\rho^{2}}cos^{2}(\theta)\Big)
\end{equation}
and the metric is
\begin{equation}\label{1metricsol}
ds^{2}=8\pi^{2}\bigg(dv_{R}^{2}+dv_{I}^{2}-\frac{\big(v_{R}dv_{I}-\bar{v}_{I}dv_{R}\big)^{2}}{\lvert v_{R}\rvert^{2}+\lvert v_{I}\rvert^{2}}\bigg)
\end{equation}
Here $\lvert\vect{q}\rvert$ is the vev of the scalar field $\phi, \ v_{R}=\rho\ \text{cos}(\theta)$ and $v_{I}=\rho\ \text{sin}(\theta)$.  
\subsection{Two U(2) Instantons}
We now move on to the case of two U(2) instantons. In the commutative case, a solution was found for the real quaternions (and with gauge group SU(2)) in \cite{Allen2012}. A solution for the noncommutative case was postulated in \cite{Iskauskas2015}, however this is not in fact correct, and an alternative solution is therefore presented here. We were unable to find a solution for the full moduli space, however we obtained a solution for the subspace defined in section \ref{CompSub}. 
 For the case of two U(2) instantons, the ADHM data has the form
\begin{equation}
\Delta=a-bx;\ a=
\begin{bmatrix}
\Lambda \\ \Omega
\end{bmatrix}=\
\begin{bmatrix}
v&w\\
\tau & \sigma^{\star}\\
\sigma&-\tau
\end{bmatrix}, \
b=\begin{bmatrix}
0&0\\
1&0\\
0&1
\end{bmatrix}
\end{equation}
Note that here $\Omega$ is constrained to be hermitian (under complex conjugation) rather than symmetric, as in the real ADHM construction. $v, w$ and $\sigma$ lie in the biquaternions, however due to the requirement that $\Omega$ be hermitian, $\tau$ remains a member of $\mathbb{H}$.  Proceeding as in section \ref{ADHM},  equation (\ref{ADHMeqn}) gives the following. First, the diagonal equations
\begin{align}\nonumber
		&v^{\dagger }v +\lvert \tau \rvert^{2} + \sigma^{\dagger}\sigma =f^{-1}_{11}\mathbb{1}+2\zeta\sigma_{3} \\
		&w^{\dagger} w +\lvert \tau \rvert^{2} + (\sigma^{\dagger}\sigma)^{\star} =f^{-1}_{11}\mathbb{1}-2\zeta\sigma_{3}
		\end{align}
		Next, the off diagonal constraints are given by
		\begin{align}\nonumber
		v^{\dagger}w+\bar{\tau}\sigma^{\star}-\sigma^{\dagger}\tau=f^{-1}_{12}\mathbb{1} \\
		w^{\dagger}v+ (\sigma^{\dagger})^{\star}\tau-\bar{\tau}\sigma=f^{-1 \star}_{12}\mathbb{1}
		\end{align}
	
This gives a total of four equations for the complex ADHM constraints. For completeness we list them here
\begin{align} \label{CADHM}\nonumber
2\imh(\bar{\sigma}_{R}\sigma_{I})-\imh(\bar{w}_{R}w_{I})+\imh(\bar{v}_{R}v_{I})=0 \\ \nonumber
\imh(\bar{w}_{R}w_{I})+\imh(\bar{v}_{R}v_{I})=-4\zeta\sigma_{3}\\ \nonumber
\imh(\bar{\tau}\sigma_{I})=\frac{ \imh(\bar{w}_{R}v_{I}+\bar{v}_{R}w_{I})}{2} \equiv \frac{\Upsilon}{2}\\  
\imh(\bar{\tau}\sigma_{R})=\frac{\imh(\bar{w}_{R}v_{R}+\bar{w}_{I}v_{I})}{2}\equiv\frac{\Lambda}{2} 
\end{align}
As a check, if we assume that our solutions to the ADHM equations are entirely real and that $\zeta=0$, we have the complex imaginary parts of all our variables being 0, and we have only the one equation which is not trivially satisfied (just as in \cite{Allen2012})
\begin{equation}
\imh(\bar{\tau}\sigma_{R})=\frac{\imh(\bar{w}_{R}v_{R})}{2}
\end{equation}
It should be noted that no new degrees of freedom are introduced compared to the real ADHM equations. Complexifying $v, w$ and $\sigma$ adds twelve degrees of freedom. However, each of the 3 new equations affecting the imaginary part of an expression adds 3 constraints, giving 9.  Recall that we have a residual O(2) symmetry on our solutions to the ADHM equation in the quaternion case, which is promoted to a  U(2) symmetry in the biquaternion case allows us to remove a further three degrees of freedom. This gives a total of 12 degrees of freedom removed, cancelling the number of new parameters and showing that there are no new solutions. We have checked this explicitly by constructing the transformation taking a commutative ADHM solution with biquaterion coordinates to the standard solution parameterised by quaternions, however this is time consuming and not especially illuminating.  \\

We were unable to solve these equations for the full biquaternion valued space, however we were able to calculate solutions restricted to the complex subspace defined in \ref{CompSub}. 
After some calculation (see appendix \ref{sec:the-noncommutative-case}), we get:
\begin{align}
\nonumber
&v_{I}=\frac{-2\zeta v_{R}\sigma_{3}}{\lvert v_{R}\rvert^{2}}\\ \nonumber
&w_{I}=\frac{-2\zeta w_{R}\sigma_{3}}{\lvert w_{R}\rvert^{2}}\\ \nonumber
&\sigma_{R}=\frac{\tau\text{Im}(\bar{w}_{R}v_{R}+\bar{w}_{I}v_{I})}{2\lvert\tau\rvert^2}=\frac{(\lvert v_{R}\rvert^{2}\lvert w_{R}\rvert^{2}+4\zeta^{2})}{2\lvert\tau\rvert^{2}\lvert v_{R}\rvert^{2}\lvert w_{R}\rvert^{2}}\tau\imc(\bar{w}_{R}v_{R})\sigma_{3}\\
&\sigma_{I}=\frac{\tau\imc(\bar{w}_{R}v_{I}+\bar{v}_{R}w_{I})}{2\lvert\tau\rvert^{2}}=-\frac{\zeta(\lvert w_{R}\rvert^{2}+\lvert v_{R}\rvert^{2})}{\lvert\tau\rvert^{2}\lvert v_{R}\rvert^{2}\lvert w_{R}\rvert^{2}}\tau\imc(\bar{w}_{R}v_{R}\sigma_{3})\sigma_{3}
\end{align}
We can check our assumption about the symmetries by checking both that our solution really does solve the ADHM equations, and that  there are no residual symmetries remaining. To show there are no residual symmetries, we consider a general U(2) transformation
\begin{equation} \label{eqn:symms2}
\Delta\mapsto
\begin{bmatrix}
1 & 0 \\
0 & R^{\dagger}
\end{bmatrix}\Delta R; \ R=
\begin{bmatrix}
a & b \\-\bar{b} & \bar{a}
\end{bmatrix}; \
a, b\in\mathbb{C}, \ \lvert a\rvert^{2}+\lvert b\rvert^{2} =1
\end{equation}
This generates the transformation
\begin{equation}\label{eq:symtrans}
\begin{bmatrix}
v & w\\
\tau & \sigma^{\star} \\
\sigma & -\tau
\end{bmatrix}\mapsto
\begin{bmatrix}
v' & w'\\
\tau' & \sigma'^{\star} \\
\sigma' & -\tau'
\end{bmatrix}\ = \
\begin{bmatrix}
av-\bar{b}w & bv+\bar{a}w \\
(\lvert a\rvert^{2}-\lvert b\rvert^{2})\tau-ab\sigma-\bar{a}\bar{b}\sigma^{\star} &
2\bar{a}b\tau-b^{2}\sigma+\bar{a}^{2}\sigma^{\star} \\
2a\bar{b}\tau+ a^{2}\sigma-\bar{b}^{2}\sigma^{\star} &
-(\lvert a\rvert^{2}-\lvert b\rvert^{2})\tau+ab\sigma+\bar{a}\bar{b}\sigma^{\star}
\end{bmatrix}
\end{equation}
Now, in appendix \ref{sec:the-noncommutative-case} we use the residual symmetry to do two things. First, we require that $\rec(\bar{\tau}\sigma)=0$. This imposes the condition that $\sigma$ has no component proportional to $\tau$. Second, we require that $\bar{w}_{R}w_{I}=\bar{v}_{R}v_{I}$. We need to work out the form of the transformation in equation (\ref{eq:symtrans}) so that the new variables $v', \tau'$ etc, also satisfy these conditions. If the form of the transformation is completely determined by this, then we know we have no remaining symmetries to consider. \\
To aid in this, we write
\begin{align}\nonumber
a &= \text{cos}\chi (\text{cos}\theta + i\text{sin}\theta ) \\ 
b &= \text{sin}\chi (\text{cos}\phi + i\text{sin}\phi)
\end{align} 
The first of these conditions, $\rec(\bar{\tau}'\sigma')=0$ requires that $\lvert a\rvert^{2}-\lvert b\rvert^{2} =0$, and hence that both $\chi=\frac{n\pi}{4}$ for $n$ from 1 to 7, and also
 $\theta=-\phi$.  This condition on $\chi$ gives the dihedral group of order 16 as a group of discrete rotations. The implications of this are discussed in \cite{Allen2012}. \\
 Now, we look at the second part of our symmetry, which is unique to the noncommutative case. Keeping $\bar{w}'_{R}w'_{I}=-\bar{v}'_{R}v'_{I}$ requires 
 $\cos(2\theta)+i\sin(2\theta)=0$. This leads to 
\begin{equation}
\theta= \frac{n\pi}{2};\ \text{e.g.} \ \theta= 0, \frac{\pi}{2}, \pi, \frac{3\pi}{2}...
\end{equation}
Now, $\theta=\frac{n\pi}{2}$ multiplies $a,b$ by $\pm 1$ and so does not change the symmetries in \cite{Allen2012}. If we set $\theta=0, \pi$ we multiply $a$ and $b$ by $\pm i$ and $\mp i$ respectively. This doesn't change $\tau$, but sends $\sigma\rightarrow -\sigma$. It also interchanges the complex real and imaginary parts of $v'$ and $w'$. Between them these two conditions fully fix the form of the transformation in equation \ref{eq:symtrans} and therefore there is no residual symmetry. \\
The next step is to calculate the scalar field, potential and metric. The calculations and results are long and not particularly illuminating, and so are given in appendices \ref{ScalarField} and \ref{MethodMetric}. Now we will go on to investigate the dynamics on the moduli space via numerical methods. 
\section{Two Instanton Dynamics}
In this section we discuss the dynamics of the instantons on the noncommutative two instanton moduli space we have constructed, solving for motion on the moduli space numerically. Where possible we use the same numerical algorithms developed in \cite{Allen2012} to produce the following figures, and for when the scattering problem becomes more complex we use a new numerical algorithm which we outline in Appendix~\ref{App:numerics}.
\subsection{The Setup}
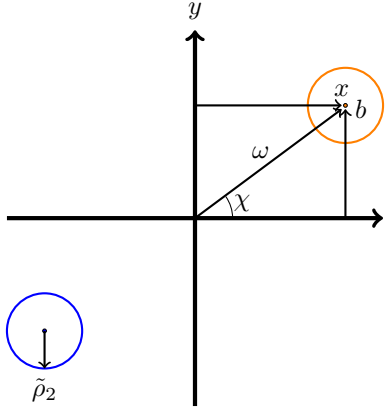
\begin{figure}
	\centering
 \begin{tikzpicture}
\draw[->,ultra thick] (-2.5,0)--(2.5,0) ;
\draw[->,ultra thick] (0,-2.5)--(0,2.5) node[above]{$y$};
\draw[orange, thick] (2,1.5) circle (0.5 cm);
\draw[blue, thick] (-2,-1.5) circle (0.5 cm);
\draw[->,thick] (0,1.5)--(1.95,1.5) node[above]{$x$};
\draw[->,thick] (2,0)--(2,1.45) node[right]{$b$};
\draw[->,thick] (0,0)--(1.95,1.45); 
\node at (0.875,0.875) {$\omega$};
\draw[fill=orange] (2,1.5) circle (0.025 cm);
\draw[fill=blue] (-2,-1.5) circle (0.025 cm);
\draw (0.5,0) arc (0:35:0.5);
\node at (0.625,0.2) {$\chi$};
\draw[->,thick] (-2,-1.5)--(-2,-2) node[below]{$\tilde{\rho}_{2}$};
 \end{tikzpicture}
 \caption{The setup of the Instantons. The instantons are located at $\pm(x,b)=\ \big(\omega\text{cos}(\chi),\omega\text{sin}(\chi)\big)$. They have size $\tilde{\rho}_{i}=\sqrt{\rho_{i}^{2}+\frac{4\zeta^{2}}{\rho_{i}^{2}}}$}
 \label{fig:paramsetup}
 \end{figure}
\begin{figure}
	\centering
	\includegraphics[width=0.45\linewidth]{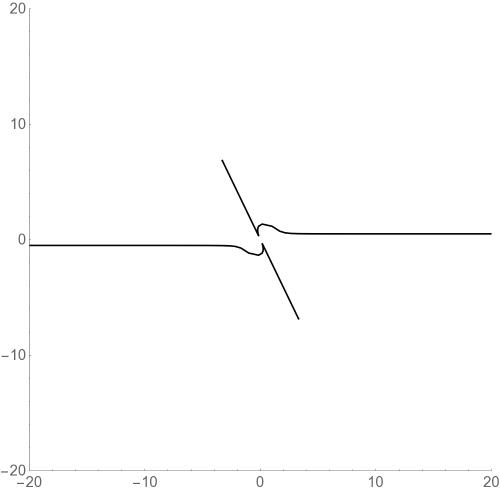} \ \ \ \ 
	\includegraphics[width=0.45\linewidth]{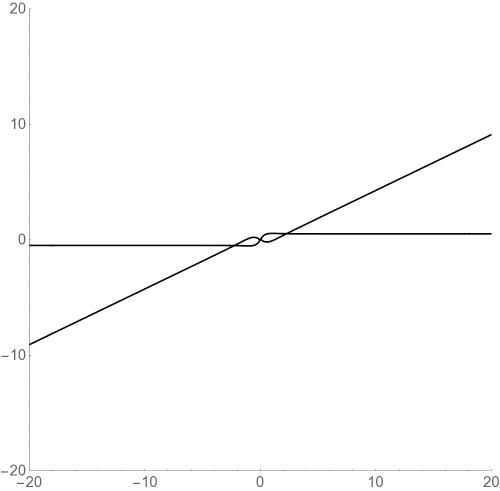}
	\caption{Scattering of Dyonic instantons with $b=0.5$ and $\zeta=1.15$. Both plots are solutions of the same initial set up, with different choices of parameterisation of the position The left plot shows the $\lvert\tau\rvert$ parametrisation, the right shows  $\pm\sqrt{\tau^{2}+\sigma^{2}}$. The radii of the instantons are not shown. In this case the $\sigma$ behaviour dominates and after the interaction the position of the instantons goes as $\frac{1}{\lvert\tau\rvert^{2}}$ Note that after 2400 time steps the $\sqrt{\tau^{2}+\sigma^{2}}$ had left the plot region, whereas the $\tau$ case has been run for 50000 time steps and still has not left the plotted area. This is because the size of the instantons (hence $v$ and $w$) becomes very large and so $\sigma$ dominates $\tau$ in the definition of the instanton position, so that merely plotting $\tau$ is a very inaccurate approximation to the position. }
	\label{taucomp}
\end{figure}

The ADHM data are in terms of $v=v_{R}+iv_{I}, w_{R}+iw_{I}, \sigma= \sigma_{R}+i\sigma_{I}$ and $\tau$.  We showed in appendix  \label{sec:the-noncommutative-case} that $v_{I}, w_{I}$ and $\sigma$ depend on the collective coordinates $v_{R}, w_{R}$ and $\tau$. As stated in section \ref{CompSub}, we are working on the subspace of the total moduli space with the collective coordinates in $\mathbb{C}\times\mathbb{C}$ rather than $\mathbb{C}\times\mathbb{H}$. This means that $v$ and $w$ are in $\mathbb{C}\times\mathbb{C}$, whilst $\tau$ is in $\mathbb{C}$ (since it lies on the diagonal, and $\Delta$ is hermitian with respect to the complex structure). \\
As is standard \cite{Manton2004}, we interpret $v,w$ as describing the embedding of the instantons into the gauge group, with $\tilde{\rho_{1}}=\lvert v \rvert$ and $\tilde{\rho_{2}}=\lvert w\rvert$ giving the physical size; and $\tau$ giving the position. This is shown in figure \ref{fig:paramsetup}. We also rewrite our independent ADHM coordinates in polar form as
\begin{align}
\nonumber v_{R}=\rho_{1}\big( \cos(\theta_{1}) +i\sin(\theta_{1})\big)\\
\nonumber w_{R}=\rho_{2}\big( \cos(\theta_{2}) +i\sin(\theta_{2})\big)\\
\tau=\omega\big(\cos(\chi)+i\sin(\chi)\big)
\end{align}
It is convenient to describe the initial conditions in cartesian coordinates $b$ and $x$ as indicated in figure \ref{fig:paramsetup}. The relation between the parameters $b, x$ and $\omega, \chi$ is
\begin{align}\nonumber
x=\omega\cos(\chi)\\ \nonumber
b=\omega\sin(\chi)\\ \nonumber
\omega=\sqrt{b^{2}+x^{2}}\\ 
\chi=\arctan(b/x)
\end{align}
Now, using the equations 
\begin{align}\nonumber
&v_{I}=\frac{-2\zeta v_{R}\sigma_{3}}{\lvert v_{R}\rvert^{2}}\\
&w_{I}=\frac{-2\zeta w_{R}\sigma_{3}}{\lvert w_{R}\rvert^{2}}\\ \nonumber
\end{align}
It follows that in the noncommutative case, the total instanton size $\rho_{i}$ is defined as
\begin{equation}
\tilde{\rho}_{i}=\sqrt{\rho_{i}^{2}+\frac{4\zeta}{\rho_{i}^{2}}}
\end{equation}
 where the index $i$ in $\rho_{i}$ is either 1 or 2, referring to the magnitude of $v$ or $w$ respectively. Calculating the scattering with general $\rho_{1}$ and $\rho_{2}$, and with general gauge embedding is very computationally expensive for the noncommutative case. Therefore we did a lot of the simulations in the, `Orthogonal' case where $\rho_{1}=\rho_{2}$ and the relative gauge angle between the two instantons is $\theta_{1}-\theta_{2}=\pi/2$.  
There are several technical issues which emerged. First of all, at the collision point there are two ingoing and two outgoing paths. Where the instanton positions nearly coincide, it is difficult to work out which ingoing and outgoing paths ought to be connected. The plotting programme we used appears to choose sensibly except in a few cases where it appears to connect the wrong pairs, as evidenced by a discontinuity in the path near the origin. This can be seen on several of the graphs, e.g. in figure \ref{peakgraph1}.\\
The second issue is with the parameterisation of the instanton position in terms of $\tau$. The position of the instanton is given by the eigenvalues of the submatrix 
\begin{equation}
\begin{bmatrix}
\tau & \sigma^{\star}\\
\sigma & -\tau
\end{bmatrix}
\end{equation}
of the ADHM data \cite{Allen2012}. Recall that
\begin{align}\nonumber
\sigma_{R}&=\frac{\text{Im}(\bar{w}_{R}v_{R}+\bar{w}_{I}v_{I})}{2}=\frac{(\lvert v_{R}\rvert^{2}\lvert w_{R}\rvert^{2}+4\zeta^{2})}{2\lvert\tau\rvert^{2}\lvert v_{R}\rvert^{2}\lvert w_{R}\rvert^{2}}\tau\text{Im}(\bar{w}_{R}v_{R})\\ 
\sigma_{I}&=\frac{\text{Im}(\bar{w}_{R}v_{I}+\bar{v}_{R}w_{I})}{2}=-\frac{\zeta(\lvert w_{R}\rvert^{2}+\lvert v_{R}\rvert^{2})}{\lvert\tau\rvert^{2}\lvert v_{R}\rvert^{2}\lvert w_{R}\rvert^{2}}\tau\text{Im}(\bar{w}_{R}v_{R}\sigma_{3})
\end{align}
In the subspace under discussion, and in the coordinates we are using, this becomes
\begin{align}\label{eqn:subspacesigmatau}\nonumber
\sigma_{R}&=\frac{i\big(\rho_{1}^{2}\rho_{2}^{2}+4\zeta^{2}\big)\big(\cos(\chi)+i\sin(\chi)\big)\sin(\theta_{1}-\theta_{2})}{2\rho_{1}\rho_{2}\omega}\\
\sigma_{I}&=\frac{-i\zeta\big(\rho_{1}^{2}+\rho_{2}^{2}\big)\big(\cos(\chi)+i\sin(\chi)\big)\cos(\theta_{1}-\theta_{2})}{\rho_{1}\rho_{2}\omega}
\end{align}
At large $\tau$, the matrix is effectively diagonal, and so the positions of the two instantons can be approximated by $\tau$ and $-\tau$ respectively. At small values of $\tau$, however, $\sigma$ becomes very large and therefore $\pm\tau$ is no longer a good description. A better approach is to diagonalise the full matrix which gives the parametrisation $\pm\sqrt{ \tau^{2}+\sigma^{2}}$ for the position (note that this is in general a complex number), though this can give a discontinuity at the origin due to the presence of the square root, with both positive and negative values. In the commutative case this is the true position of the instanton. In the noncommutative case, the noncommutativity of the underlying space means that the meaning of, `true position' is not clear, however it still makes more sense to use the $\sqrt{\tau^{2}+\sigma^{2}}$ parametrisation, as the presence of $\zeta$ in $\sigma$ means that this becomes more important (as shown in figure \ref{taucomp}).
We now move on to looking at the graphs. 
\subsection{ Pure Instantons }\label{sec:fourparamspace}
\begin{figure}
	\includegraphics[width=0.8\linewidth]{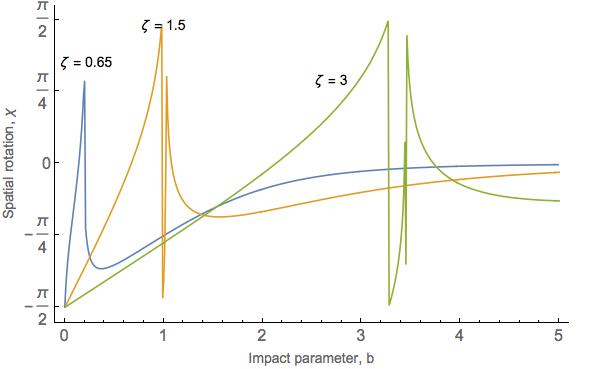}
	\caption{Plot of scattering angle vs. impact parameter for different values of $\zeta$. From left to right we have $\zeta=\{0.65,1.5,3 \}$}
	\label{ScavsImpMultGraph}
\end{figure}

\begin{figure}
\includegraphics[width=0.6\linewidth]{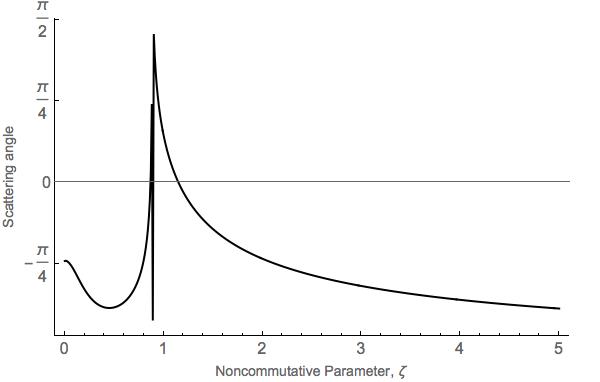}
	\includegraphics[width=0.6\linewidth]{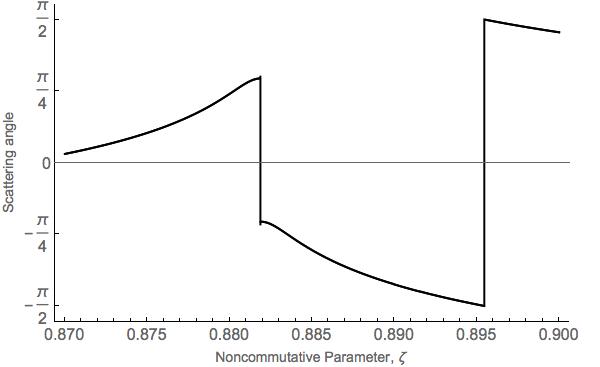}
\caption{Change of scattering angle (left) with noncommutative parameter $\zeta$ for $b=0.5$, with the other parameters as discussed in the main body of the text. On the right is shown the area around the discontinuity, which seems to be a region where the numerics have confused $\pm\theta$.    }\centering
\label{scaoverallgraph}
 \end{figure}

\begin{figure}
	\includegraphics[width=0.4\linewidth]{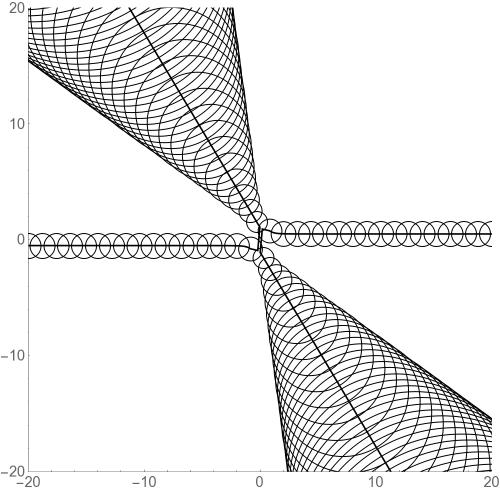} \ \ \ \
	\includegraphics[width=0.4\linewidth]{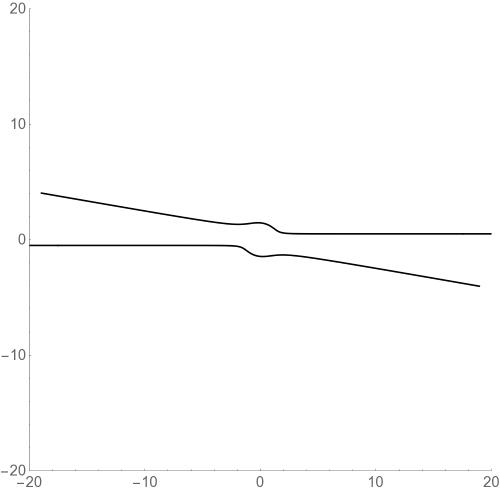}\\
	\includegraphics[width=0.4\linewidth]{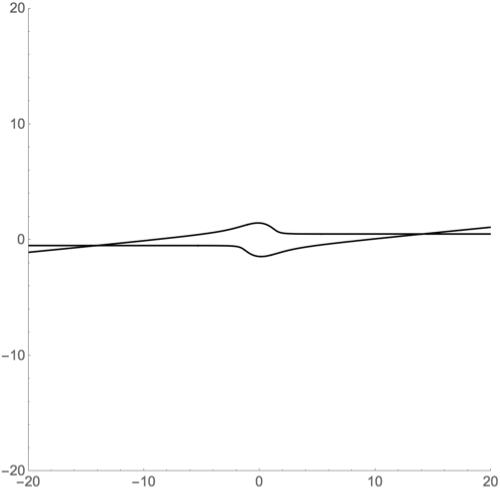}\ \ \ \ \
		\includegraphics[width=0.4\linewidth]{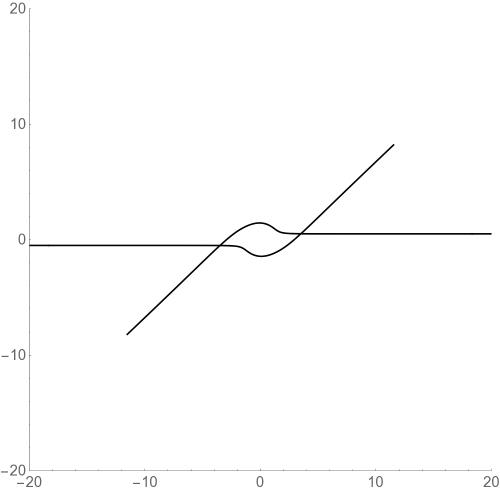}				
	\caption{Scattering for two instantons with $b=0.5$, and, moving in each row from left to right,  $\zeta=\{0.1,0.86, 0.87,0.88\}$; This corresponds to the region to the left of and around  the peak in figure \ref{scaoverallgraph}. Where the sizes are not shown this is in order to make the trajectories clearer. Note that the instantons go from glancing off one another, to moving parallel, to crossing over, and then deflecting }
	\label{Sca1Graph}
\end{figure}

\begin{figure}
	\includegraphics[width=0.3\linewidth]{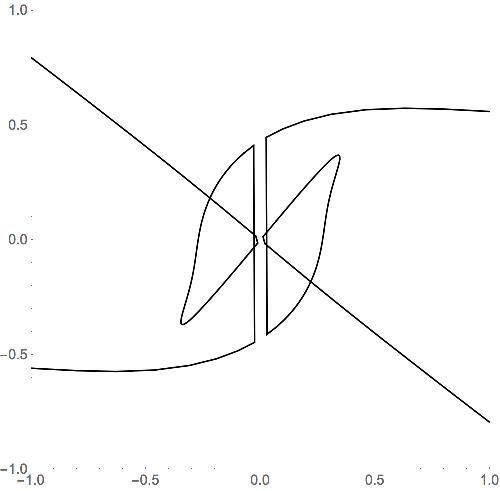} \ \ \ \ 
	\includegraphics[width=0.3\linewidth]{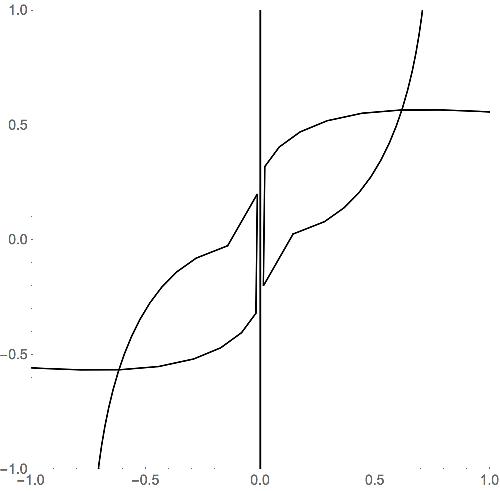}\ \ \ \ \ 
	\includegraphics[width=0.3\linewidth]{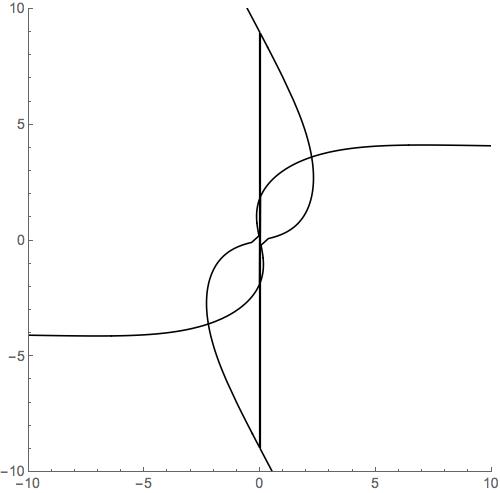}
	\caption{Graphs of the interaction around the peak, zoomed in at the center, with $\zeta=\{0.8187, 0.89, 0.9\}$. The vertical lines on the graph are the results of confusion about which parts of the trajectory belong to which instanton, and can be ignored. More detailed interpretation of the graphs is given in the main text.}
	\label{peakgraph1}
\end{figure}

\begin{figure}
	\includegraphics[width=0.4\linewidth]{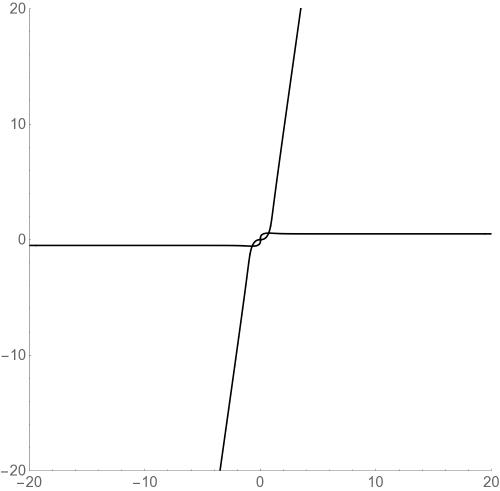}\
	\includegraphics[width=0.4\linewidth]{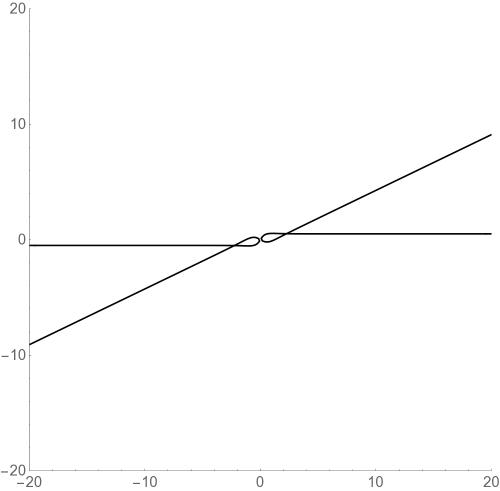}\ \
	\includegraphics[width=0.4\linewidth]{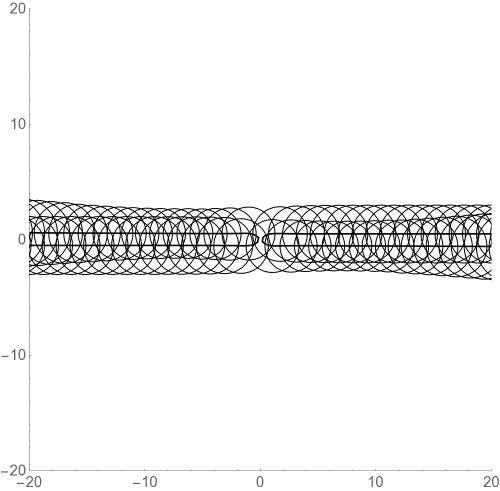}\ \
	\includegraphics[width=0.4\linewidth]{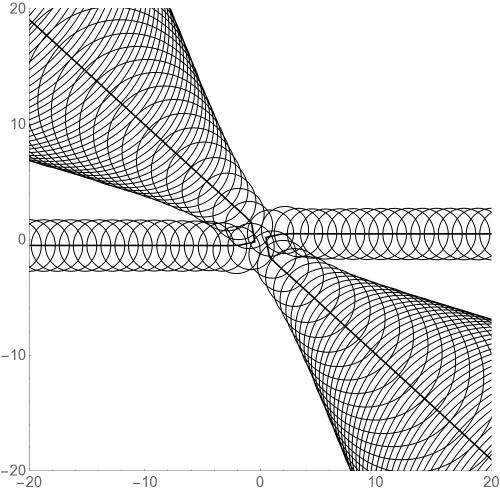}
	\caption{Scattering for two instatons with $b=0.5$, and $\zeta={0.9, 1 ,1.15 ,2}$. This corresponds to the right side of the peak. Note that the instanton angle begins to turn back on itself, until the scattering becomes a direct repulsion $\zeta=1.15$, then opens to about $\pi/4$}
	\label{Sca2graph}
\end{figure}

We start with the four parameter orthogonal instantons. There are two procedures we can use to investigate the moduli space dynamics when noncommutativity is introduced. The first is to start off with a particular example of commutative scattering and see how turning on the noncommutativity affects this. The second relies on the fact that choosing a value for the noncommutative parameter $\zeta$ sets an overall scale. We can therefore `scan' the parameter space for interesting behaviour by varying one of the other parameters at a time for a fixed value of $\zeta$. It should be noted that this scanning process is not in itself sensitive to periodic ambiguities in the scattering angles -- e.g. instantons moving parallel and not interacting and instantons reflecting directly off each other would both register a scattering angle of zero. Therefore we must supplement this scanning by looking at individual plots to check the interpretation of the scattering angles we have found.  \\
Both these methods have their uses, and we will use each in turn. With the first method, we will see how a typical example of scattering changes with the parameter $\zeta$. We will also investigate what happens to the orthogonal scattering in the noncommutative case. We will then use the second method to see if there is interesting systematic behaviour associated with the other ADHM parameters. \\
  First, we take a typical example of scattering in the commutative case, and see what happens when we add in noncommutativity. In this case, the parameters $\{ \rho, \theta, b, x\}$ take the values $\{1,0,0.5,50\}$ and their initial derivatives are $\{0,0,0,-0.03\}$. The change in scattering angle as we change the value of $\zeta$ from 0 to 5 is shown in figure \ref{scaoverallgraph}.  The most obvious feature  is the presence of a peak. This is a general feature of scattering as we change $\zeta$ -- see figure \ref{ScavsImpMultGraph}.
 The peak is hard to resolve numerically -- there seems to be a discontinuity. To analyse this we can zoom in on that section of the graph (figure \ref{peakgraph1}). Part of the issue seems to be a numerical error based on the code jumping between $\pm\theta$ at $\pi/4$ and $\pi/2$. This cannot completely explain the phenomenon however.  Looking at the graphs in \ref{peakgraph1}, the instantons appear to merge then divide again in a way that, `swaps over' the ingoing and outgoing tracks. This may indicate the presence of a bound state at the cusp, which the numerics cannot fully resolve. Another possibility is that there is suddenly a second scattering at right angles. It is not clear why there should be such a sharp transition, but perhaps it happens when the instantons become large enough to overlap during the scattering process. This would be a good topic for future work with more powerful numerical methods.  \\
 In general, the effect of the noncommutativity is to increase the repulsion between the instantons. Initially, the instanton scattering angle seems to rotate anticlockwise, going from glancing off each other, to moving parallel, to crossing over. This occurs rapidly as $\zeta$ changes from 0.85 until about 0.88 (figure \ref{Sca1Graph}).  \\
 At the first apparent discontinuity, the instantons change from moving across each others paths, to repelling and turning back on themselves, so that their paths form a loop near the interaction point. This change happens somewhere between $\zeta=0.8818695$ and $\zeta=0.88187$.  The beginnings of the looping behaviour can be seen in the first graph in figure \ref{peakgraph1}, however as can be seen in the second graph there is no way of assigning the trajectories to different instantons. This is a general feature of instanton dynamics- you cannot distinctly seperate instantons when they get too close \cite{Manton2004}. \\
  Some of the issue with correctly defining the angle can be seen from the bottom two graphs in figure \ref{peakgraph1}, where the loops are joined in two different ways. Comparing the last graph in \ref{peakgraph1}, the last graph in figure \ref{scaoverallgraph} and the first graph in figure \ref{Sca2graph} indicates that the particles seem to loop back on themselves. This is further evidence for the fact that discontinuities in the scattering angle graph \ref{peakgraph1}  are based on a breakdown of the notion of the instanton positions as they begin to intersect.   \\
 After the peak (figure \ref{Sca2graph}), the scattering angle appears to rotate clockwise -- the loop at the interaction point is, `unwound'. This leads to them then repelling entirely before the angle widens to about $\pi/4$, with the instantons repelling rather than glancing off each other as they did at the start.  


We observed this behaviour for a number of different set ups, including where the instantons began too far apart to originally interact. 
An overall feature of all this graphs is that whereas in the commutative case the instantons shrink through zero size then expand again, in the noncommutative case, as we would expect, they shrink to a finite size before expanding; since due to the noncommutativity the zero size point cannot be reached. \\

\begin{figure}
	\includegraphics[width=0.4\linewidth]{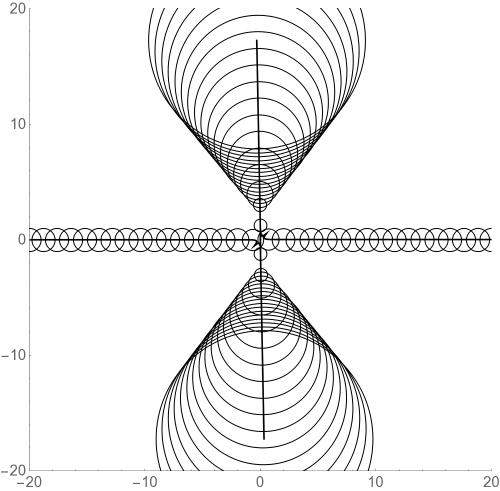}
	\includegraphics[width=0.4\linewidth]{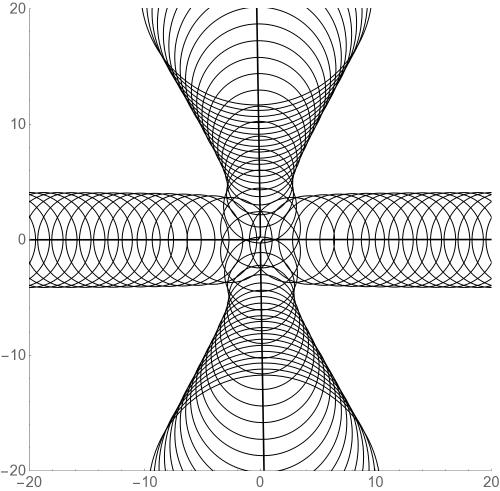}
	\caption{Examples of scattering behaviour for the orthogonal scattering with $b=0$ for $\zeta=\{0,2\}$.}
	\label{sca1perp}
\end{figure}

Plotting the scattering angle for differing values of the impact parameter $b$ shows the same distinctive spike for a particular value of $b$ (figure \ref{ScavsImpMultGraph}). As a numerical check, if we interchange the roles of $b$ and $\zeta$ by plotting the scattering angle for varying $\zeta$ whilst keeping $b$ fixed,  the spike appears at the same $(b, \zeta)$ coordinates.  If we plot the graphs of scattering angle vs. impact parameter for different values of $\zeta$ we see that the overall behaviour stays the same, however the position of the peak moves to the right as $\zeta$ increases as shown in figure \ref{ScavsImpMultGraph}. This behaviour only appears when $\zeta\neq0$, and therefore seems to be unique to the noncommutative case. This would be another topic for further investigation. 
\\The next case we will look at is the case where the scattering is orthogonal in the commutative case (so $b=0$). This remains consistently orthogonal in the noncommutative case -- see figure \ref{sca1perp}. Overall the scattering keeps its perpendicular character. We can explain this analytically in a similar way as in the commutative case in \cite{Allen2012}. As discussed above, the location of the instantons is described by a combination of $\tau$ and $\sigma$. Because $\sigma$ behaves like $1/\lvert\tau\rvert$, the change in which parameter dominates happens when $\lvert\tau\rvert=\lvert\sigma\rvert$.  Recall the definitions of $\tau$ and $\sigma$ in equation (\ref{eqn:subspacesigmatau}). For the case of orthogonal scattering, $\chi=0$. Therefore $\tau=\omega$, and so lies entirely on the $x$-axis. On the contrary, $\sigma$ is proportional to $i$ and so lies on the $y$-axis. Therefore, as the dominant parameter in the position changes between $\tau$ and $\sigma$, the instanton motion changes from the $x$-axis to the $y$-axis and so they scatter orthogonally. \\

\begin{figure}
	\includegraphics[width=0.4\linewidth]{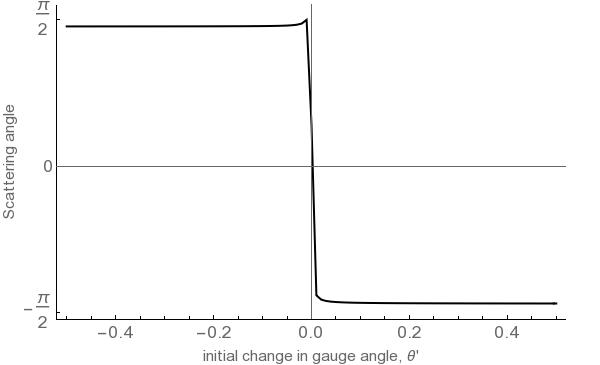}
\includegraphics[width=0.4\linewidth]{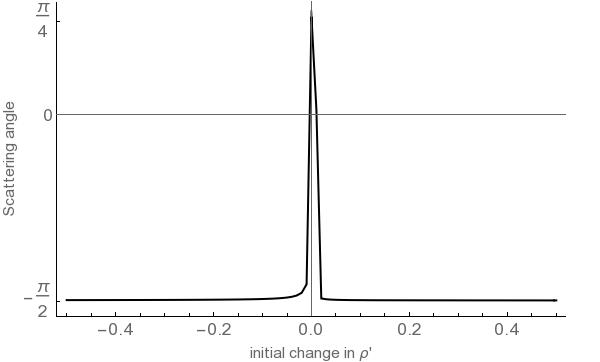}
	\caption{(left) Graph showing variation of scattering angle vs. gauge angle $\theta$ for $b=0.5$ and $\zeta=1$. Note the jump in scattering angle between positive and negative $\dot{\theta}$. (right) Graph showing variation of scattering angle vs. $\dot{\rho}$ for $b=0.5$ and $\zeta=0.1$. Note that here there is no jump in the scattering angle, unlike in the case of $\dot{\theta}$}
\label{fig:thetaprvaroverall}
\end{figure}

\begin{figure}
	\centering
	\includegraphics[width=0.45\linewidth]{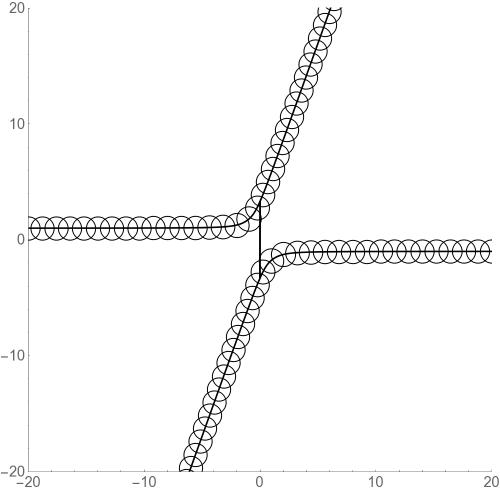} \ \ \ \
	\includegraphics[width=0.45\linewidth]{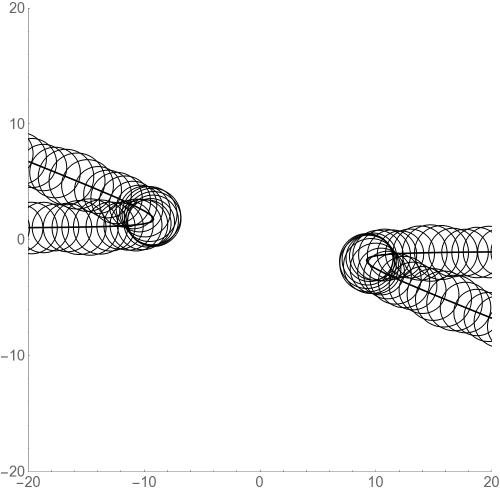}
	\caption{Plot of dyonic instanton scattering for $b=-1, \ \lvert q\rvert=0.1,\ \zeta=0$ (above), $ \zeta=1$ (below). Note the visible oscillations on the right hand graph.}
	\label{Dyongraph2}
\end{figure}

\begin{figure}
	\centering
	\includegraphics[width=0.6\linewidth]{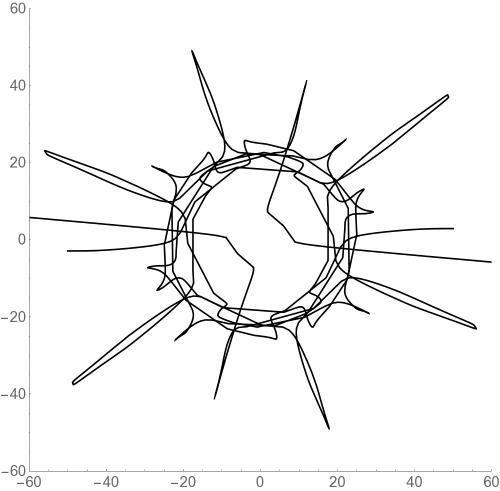}
	\caption{Graph of scattering with orbiting behaviour, with $\zeta=0.5, \ b=0.5, \ q=0.00438$}
		\label{fig:qvarorbit}
\end{figure}

We then used the scanning method to look for interesting behaviour amongst the other parameters. Fixing $\zeta$ fixes the length scale of the system, therefore we investigated the behaviour of the system keeping $\zeta$ at a constant value of 1, and looking at how the scattering angle of the instantons depends on the other parameters. 
The variables for which there was notable behaviour were  $\dot{\rho}$ and $\dot{\theta}$. These showed similarly interesting behaviour in both cases, and so I will discuss them together. As can be seen in figure \ref{fig:thetaprvaroverall}, in both cases, a small pertubation in $\dot{\rho}$ and $\dot{\theta}$ causes almost orthogonal scattering, no matter what the initial scattering angle. The difference is that, as shown in figure \ref{fig:thprsca1}, there seems to be jump of the scattering angle from positive to negative,  for positive and negative $\dot{\theta}$, which is not present for $\dot{\rho}$. However it is unclear even from individual scattering graphs if this is a misidentification of incoming and outgoing particles. Even if there is no scattering in the, `base' case where both are zero, as in figure \ref{sca1perp}, we still get the same orthogonal scattering behaviour, however there is not the same jump in scattering angle at the origin. The reason for this  behaviour seems to be that changing either of these parameters from zero makes the instanton size very large, causing a high degree of interaction (and hence orthogonal scattering) no matter what the initial separation is. There is a subtlety in that $\dot{\rho}$ is not the variation in the actual size, but only in the parameter $\rho$. The variation in the actual size is give by 
\begin{equation}
\dot{\tilde{\rho}}= \frac{\dot{\rho}\rho-\frac{4\zeta^{2}\dot{\rho}}{\rho^{3}}}{\sqrt{\rho^{2}+\frac{4\zeta^{2}}{\rho^2}}}
\end{equation} 
Therefore $\dot{\rho}$ is a good approximation when $2\zeta/\rho$ is small. We have been careful to only consider such cases
We now move on to dyonic instantons. We follow the same method as before. First we look at how changing $\zeta$ changes a specific case of scattering. Then we use the scanning technique to look for interesting behaviour linked to varying the ADHM parameters. In the dyonic case, the length scale is still set by $\zeta$, however the scale of the time dimension is no longer arbitrary, but is set by $\lvert q\rvert$. Therefore we must consider the consequences of varying both.\\
Looking at specific scattering examples, we again see that the noncommutative parameter initially introduces a repulsive effect (eg. figure \ref{Dyongraph2}). Here, the basic values of the parameters are as in the pure case, except that we give $\theta$ a small initial velocity of $0.1$ in order to avoid numerical issues. Any changes to these parameters will be discussed in the captions to the graphs. As discussed in \cite{Allen2012}, dyonic instantons oscillate along their motion, and this effect is much more observable with the noncommutativity turned on. Scanning along the scattering angle in both the commutative and noncommutative case gave very noisy graphs from which it is hard to deduce any global behaviour. However we found some interesting examples of orbiting behaviour, especially for small $q$ relative to the other parameters -- a particularly impressive example is figure \ref{fig:qvarorbit}. \\

\begin{figure}
	\includegraphics[width=0.5\linewidth]{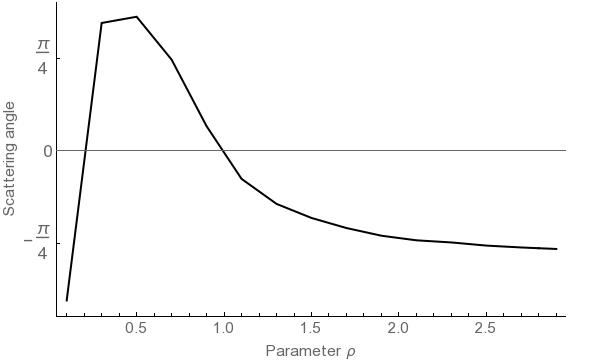}
	\ \ \ \ 
		\includegraphics[width=0.5\linewidth]{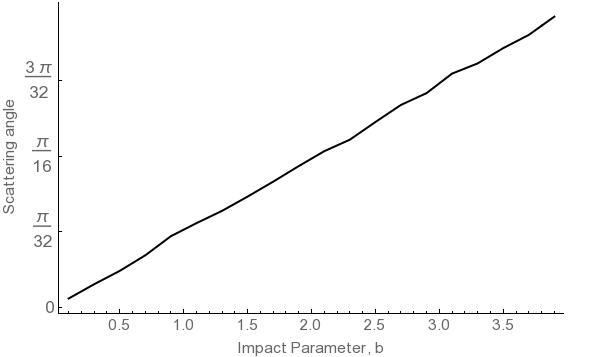}
	\caption{Left: Scattering angle vs. parameter $\rho$ for $\zeta=q=0.1$. As discussed in the text, we expect $\rho$ to be a good approximation to the true initial size after the peak
	Right: Scattering angle vs. Impact parameter for $\zeta=q=0.5$}. The case b=0 is discussed in figure \ref{sca1perp}, and not included in this graph. 
	\label{fig:dyrhovar}
\end{figure}

We can then systematically look for interesting behaviour amongst the remaining parameters. A recurring feature was the presence of stable combinations of $\zeta$ and $q$ for which there was a clear pattern of behaviour, with no observable pattern outside of these regions.   \\
We started by looking at varying $\theta$, but this did not yield any interesting systematic behaviour -- only random noise. We then looked at $\rho$. Here there seemed to be a window where the behaviour matched the commutative case, e.g. with $\zeta=0.1$ and $q=0.1$, as shown in figure \ref{fig:dyrhovar}. Exploring around that point showed that the behaviour persisted with roughly $\zeta<1$ and with $q>0.08$ \\
We then moved on to looking at $b$. Here there was similar behaviour. At low $\zeta$ there did not seem to be any overall pattern, however increasing $\zeta$ led to graphs having a linear pattern, as in figure \ref{fig:dyrhovar}. This seems to be two examples of the same phenomenon, with very different scaling caused by the differences in the relative values of $\zeta$. The magnitude of $q$ did not seem to have a major effect on whether any linear pattern in the behaviour was observed past a certain point, but continuing to make $\zeta$ larger caused the non linear behaviour to return.  \\
Finally, we did not find any discernible patterns for $\dot{\theta}$ or $\dot{\rho}$ either. This difference as compared to the pure case is probably because the potential prevents the instanton size from growing large in the dyonic case, and therefore the transition to orthogonal scattering cannot occur.

\subsection{The Six Parameter Space}\label{sec:sixparamspace}
\begin{figure}
	\includegraphics[width= 0.55\linewidth]{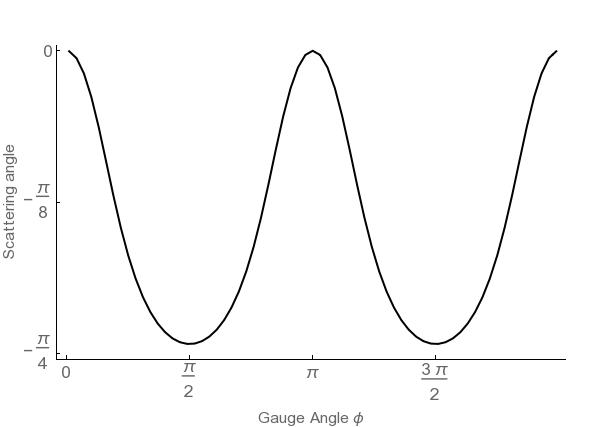}\ \ \ \ 
		\includegraphics[width= 0.55\linewidth]{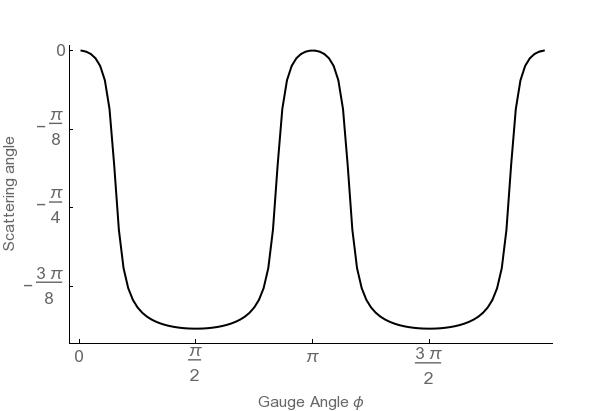}
	\caption{Left: Graph of varying scattering angle $\phi$ with $\zeta=0$, $\rho_{1}=\rho_{2}=1$ and $b=0.5$.
	Right: Graph of varying scattering angle $\phi$ with $\zeta=0$, $\rho_{1}=\rho_{2}=1$ and $b=0.1$.}
	\label{fig:6paramcommoverall1}
\end{figure}

\begin{figure}
	\includegraphics[width= 0.55\linewidth]{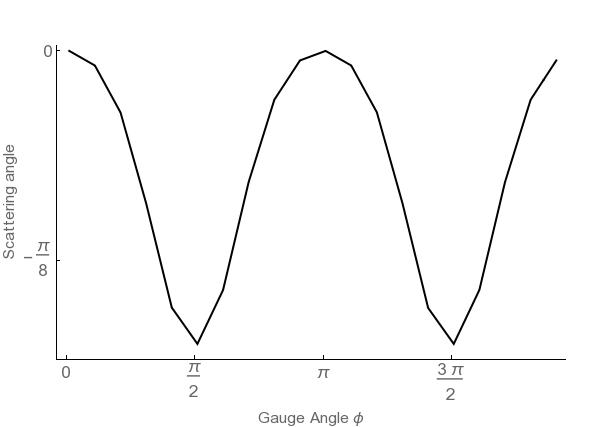}\ \ \ \ 
	\includegraphics[width= 0.55\linewidth]{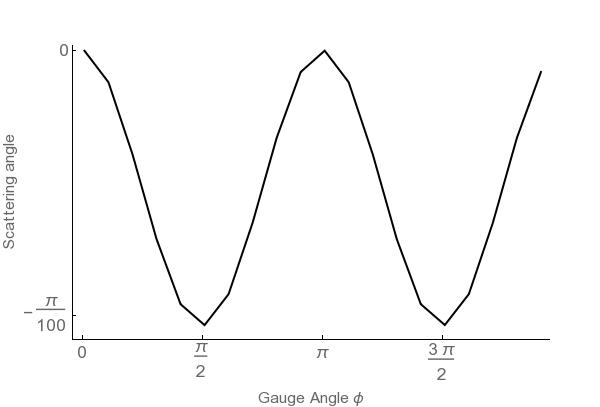}
	\caption{Graph showing variation of scattering with gauge angle $\phi$, where $\zeta=0$ and $b=0.5$. In both graphs $\rho_{1}=1$. In the left graph, $\rho_{2}=5$, and in the right graph $\rho_{2}=0.1$. Note that the scattering angle is much smaller in this case.}
	\label{fig:6paramcommoverall34}
\end{figure}

\begin{figure}
	\includegraphics[width= 0.5\linewidth]{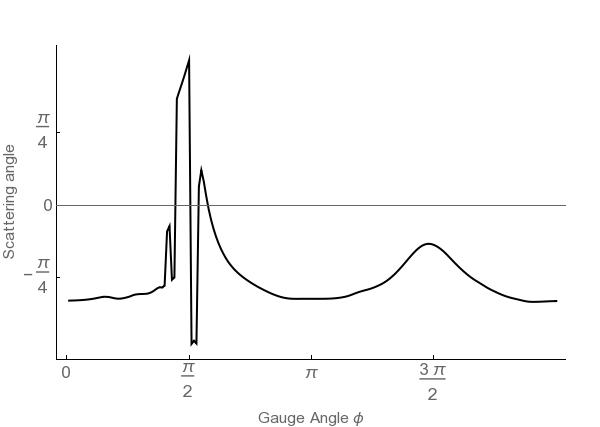} \ \ \ \ 
	\includegraphics[width= 0.5\linewidth]{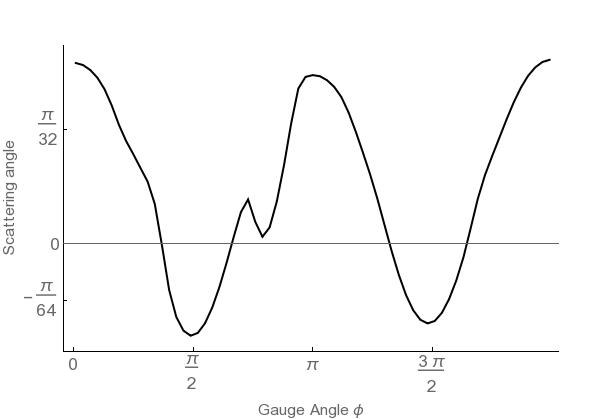}
	\caption{Left: Graph showing variation of scattering angle $\phi$ with $\zeta=1$, $\rho_{1}=\rho_{2}=1$ and $b=0.5$. The true instanton size is therefore $\sqrt{2}$, and so is roughly comparable to the separation. The splitting of the left peak appears to be a numerical error.
	Right: Graph showing variation of scattering angle $\phi$ with $\zeta=1$, $\rho_{1}=\rho_{2}=1$ and $b=4$. The true instanton size is therefore $\sqrt{2}$, and so is much smaller than the separation. Note that the sinusoidal form is much more preserved, but now oscillates around zero rather than away from it }
	\label{fig:6paramoverall1}
\end{figure}

\begin{figure}
	\includegraphics[width= 0.5\linewidth]{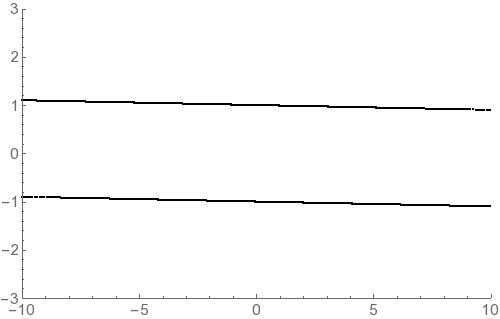}
	\includegraphics[width= 0.5\linewidth]{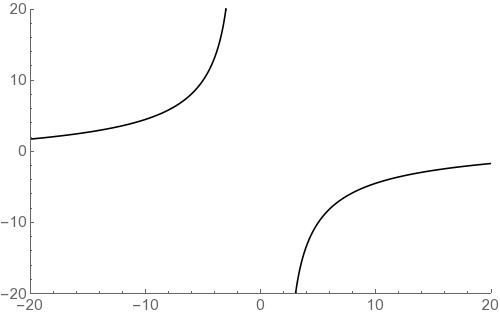}
	\caption{Graph showing scattering examples from figure \ref{fig:6paramoverall1}, with $\phi=\pi$ left and $\phi=3\pi/2$ right. Note that the scales are different on the two graphs, and that the behaviour is extremely different. }
	\label{fig:6paracommsca1}
\end{figure}

\begin{figure}
	\centering
	\includegraphics[width= 0.65\linewidth]{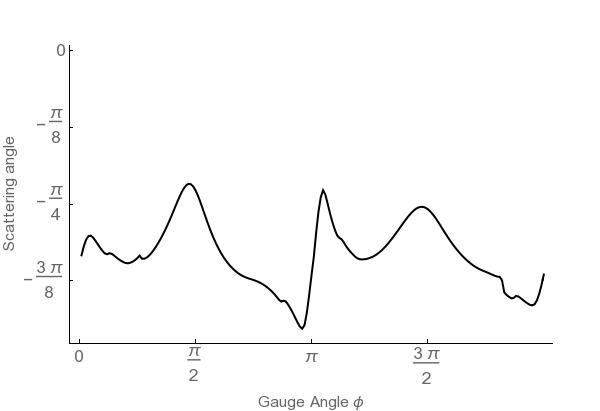}
	\caption{Graph showing varying gauge angle $\phi$ with $\zeta=1$, $\rho_{1}=1; \rho_{2}=5$ and $b=0.5$. The true instanton sizes are $\sqrt{2}$ and just over 25 respectively.}
	\label{fig:6paramoverall3}
\end{figure}

We now move on to look at the full six parameter space, where the instantons are free to have different sizes and to vary in their gauge angle. The additional parameters add greatly to the complexity of the numerics, which we describe in Appendix \ref{App:numerics}. Due to this it is no longer possible to use the numerical algorithm from \cite{Allen2012} as we had in the four parameter case and so we had to use a new algorithm, outlined in Appendix \ref{App:numerics} to produce the figures in this section.\\
There are two parameters to examine here. These are the relative gauge angle $\phi$ and the relative sizes of the instantons. Unless otherwise stated, the initial conditions for $\{\rho_{1}, \rho_{2}, \theta, x\}$ take the values $\{1,1,0,50\}$, and the initial derivatives of all parameters are zero, except $\dot{x}=-0.03$. In the noncommutative case it is tricky to systematically explore the latter as the instanton sizes are nonlinear functions of $\zeta$ and the $\rho_{i}$. Therefore we chose to keep $\zeta$ fixed to set the overall length scale, and to vary the impact parameter rather than the instanton size, looking at cases where the separation was much smaller than, larger than and of the same order as the sizes of the instantons. Initially we kept the instantons the same size. We then checked the behaviour in three cases $\rho_{1}<b<\rho_{2}$, $b<\rho_{1}<\rho_{2}$ and $\rho_{1}<\rho_{2}<b$. \\
In the commutative case, varying the gauge angle produces a clear sinusoidal variation (figure \ref{fig:6paramcommoverall1}). This pattern held for different values of the impact parameter, however when the impact parameter was small compared to the instanton size, the varation takes on more of a, `square' shape (figure \ref{fig:6paramcommoverall1}). As can be seen both from these two figures and from the scattering angles in figure \ref{fig:6paracommsca1}, at $\phi=n\pi$, where the instantons are parallel in the gauge group, the interaction between the instantons disappears and they just move past one another. Conversely, the instantons interact most strongly at $\phi=n\pi+\pi/2$, where they are orthogonal in the gauge group. Changing the relative sizes of the instantons did not seem to affect this sinusoidal behaviour, but it did change the strength of the interaction, with the scattering angle decreasing when the Instantons were different sizes, with smaller sizes making the scattering angle smaller (figure \ref{fig:6paramcommoverall34}).\\
The behaviour in the noncommutative case is not so simple. The outline of the sinusoidal pattern is still present, but it is significantly disrupted, as in figure $\ref{fig:6paramoverall1}$. Increasing the impact parameter somewhat restores the behaviour (figure $\ref{fig:6paramoverall1}$). There is therefore much less variation in the scattering angle for the noncommutative case. The Instantons also no longer stop interacting when they are parallel in the gauge group, instead oscillating between minimum and maximum scattering angles, as in figure \ref{fig:6paramoverall1}. Making one of the instantons smaller than the other and the impact parameter did not seem to have too much of an effect, however, making one larger than the impact parameter further disrupted the sinusoidal pattern, as in figure $\ref{fig:6paramoverall3}$. 
\subsection{Conclusions}
We end this section by reviewing the main results. We looked at both the full six parameter space, and also a four parameter subspace where the instantons were orthogonally embedded in the gauge group. This was necessary to analyse the dyonic case. Overall, increasing the noncommutative parameter $\zeta$ increases the repulsion between the instantons. The form this takes is not straightforward, and in the pure instanton case involves a peak with strange behaviour which requires a future, more detailed analysis with more sophisticated simulations. However in general even if the instantons begin by not interacting, they move from glancing off each other, to reflecting entirely as the parameter $\zeta$ increases. \\
We also found that orthogonal scattering was present in the noncommutative case as well as the commutative case.
Systematically looking at the other parameters, we saw that, as expected, increasing $\rho$ strengthens the repulsive effect of the scattering, and increasing the separation $b$ decreases it. Further interesting behaviour was observed seeing how the scattering changed when the quantities $\dot{\rho}$ and $\dot{\theta}$ were varied. For any nonzero value of these initial velocities, the scattering rapidly became almost orthogonal. This seems to be because making these parameters nonzero causes a rapid increase in the instanton size, and hence a very strong interaction. \\
This behaviour is not found in the dyonic case; probably because the presence of the potential suppresses the instanton size. In the dyonic case there was the additional feature of orbiting behaviour, some of a high winding number and great complexity. \\
Finally, we were able to use the six parameter pure instanton case to analyse changing the gauge embedding. In general we found that the scattering oscillated with the gauge angle, but that this was suppressed as $\zeta$ was increased. 
  \section{Three Instantons}
 We now move on to the case of three instantons in SU(2) Yang Mills. Here we only consider a commutative background, not a noncommutative one. We also use the usual version of the commutative ADHM construction with the quaternions rather than the biquaternion construction outlined above. As before, we begin by solving the ADHM constraints. We then calculate the scalar field for the dyonic case and use this to calculate the moduli space potential. Finally, we calculate the moduli space metric. The results in this section are original. There is, however, some related work in \cite{ThesisCochburn} and its related papers. There, some three monopole solutions are found, using two methods involving writing the solution as a reduction of the ADHM equations. The first method is to use the JNR ansatz. This corresponds to taking $\Omega$ to be diagonal in our notation. The second is to calculate Axial monopoles using ADHM data which has axial symmetry imposed on it via the Manton- Sutcliffe method. In our notation, there is a non-diagonal but specific form for $\Omega$, and the instanton size $v$ is chosen to be zero. These specific symmetries do not seem to match the ones we have chosen, and hence it it not immediately clear how the results in that work relate to those presented here, but it would be interesting and worthwhile to pursue this in future. \\ 
Since we are in the commutative case, we need to solve the equation $\Delta^{\dagger}\Delta=0$. For three instantons in SU(2) Yang Mills, the ADHM data $\Delta$ is
\begin{equation}
\begin{bmatrix}
\Lambda\\
\Omega
\end{bmatrix}
= \
\begin{bmatrix}
u & v & w\\
\tau_{1} & \sigma_{1} & \sigma_{2}\\
\sigma_{1} & \tau_{2} & \sigma_{3}\\
\sigma_{2} & \sigma_{3} & \tau_{3}
\end{bmatrix}
\end{equation}
where the entries of $\Delta$ all lie in $\mathbb{H}$. With $\Delta$ as given above, we have three equations, one for each basis vector of $o(3)$. These are
\begin{align}\nonumber
&\imh\big(\bar{u}v+(\bar{\tau}_{1}-\bar{\tau}_{2})\sigma_{1}+\bar{\sigma}_{2}\sigma_{3}\big)=0\\ \nonumber &
\imh\big(\bar{u}w+(\bar{\tau}_{1}-\bar{\tau}_{3})\sigma_{2}+\bar{\sigma}_{1}\sigma_{3}\big)=0\\  &
\imh\big(\bar{v}w+(\bar{\tau}_{2}-\bar{\tau}_{3})\sigma_{3}+\bar{\sigma}_{1}\sigma_{2}\big)=0
\end{align}
Note that these are now nonlinear in the ADHM data.  Again, we were unable to find a solution on the full quaternion moduli space; however, if we restrict to the complex subspace as in the noncommutative 2 instanton case, and use the three residual symmetries to set the real parts of the $\sigma_{i}$ to zero, the terms in $\text{Im}(\bar{\sigma}_{i}\sigma_{j})$ vanish, and  we can solve as
\begin{align}\label{eqn:3sigma}\nonumber
&\sigma_{1}=\frac{\tau_{1}-\tau_{2}}{\lvert\tau_{1}-\tau_{2}\lvert^{2}}\bigg(\alpha-\imc(\bar{u}v)\bigg)\\ \nonumber 
&\sigma_{2}=\frac{\tau_{1}-\tau_{3}}{\lvert\tau_{1}-\tau_{3}\lvert^{2}}\bigg(\beta-\imc(\bar{u}w)\bigg)\\ 
&\sigma_{1}=\frac{\tau_{2}-\tau_{3}}{\lvert\tau_{2}-\tau_{3}\lvert^{2}}\bigg(\gamma-\imc(\bar{v}w)\bigg)
\end{align}
For constants $\alpha, \beta, \gamma \in\mathbb{R}$. These are then constrained by the condition $\rec(\sigma_{i})=0$ to be
\begin{align}\nonumber
&\alpha=-\frac{\imc(\tau_{1}-\tau_{2})\imc(\bar{u}v)}{\text{Re}(\tau_{1}-\tau_{2})}\\ \nonumber
&\beta=-\frac{\imc(\tau_{1}-\tau_{3})\imc(\bar{u}w)}{\text{Re}(\tau_{1}-\tau_{3})}\\ 
&\gamma=-\frac{\imc(\tau_{2}-\tau_{3})\imc(\bar{v}w)}{\text{Re}(\tau_{2}-\tau_{3})}
\end{align}
The minus sign comes from the fact that each $\imc$ comes with a $\sigma_{3}$, which multiply together to give $-1$. The above equations give a solution for the complex subspace. As with the two instanton case, the solutions for the scalar field, metric and potential are long and are given in Appendix \ref{App:ThreeInst}

\subsection{3 Instanton Dynamics} \label{sec:threeinst}
\begin{figure}
	\includegraphics[width=0.4\linewidth]{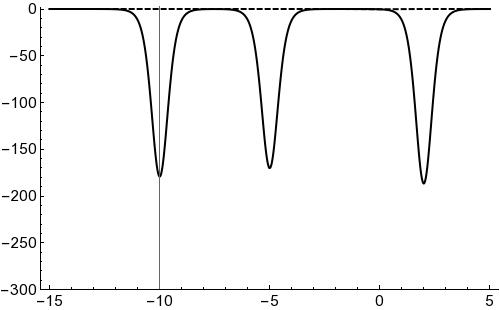}
	\includegraphics[width=0.4\linewidth]{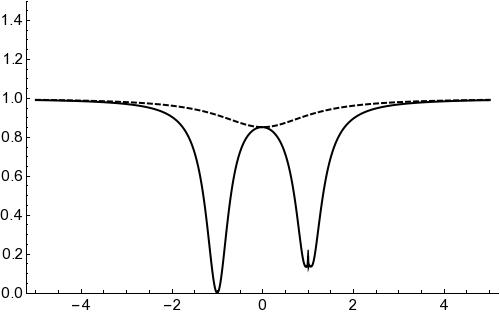}
	\caption{(left) Plot of the scalar field profile for three separated instantons. (right)Plot of the scalar field profile for two instatons, with the third far separated off to the right}
	\label{Sc3Sep}
\end{figure}

\begin{figure}
	\includegraphics[width=0.3\linewidth]{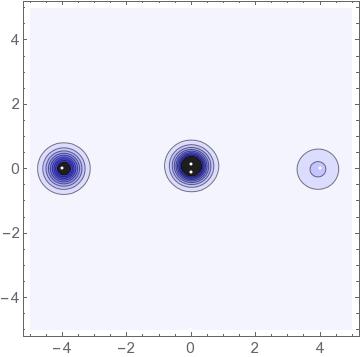}
	\includegraphics[width=0.3\linewidth]{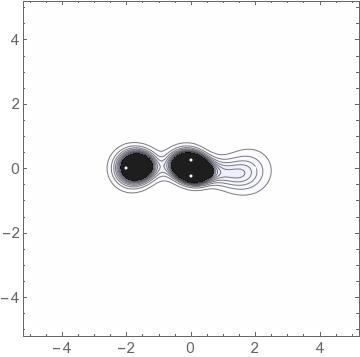}
		\includegraphics[width=0.3\linewidth]{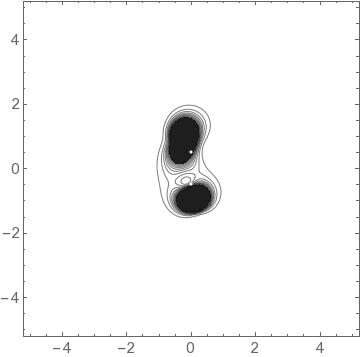}
		\includegraphics[width=0.3\linewidth]{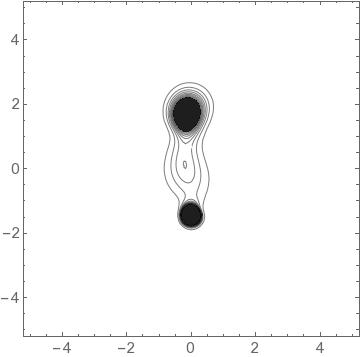}
			\includegraphics[width=0.3\linewidth]{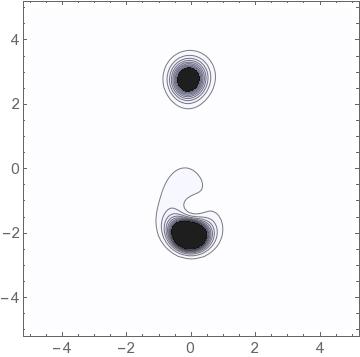}
				\includegraphics[width=0.3\linewidth]{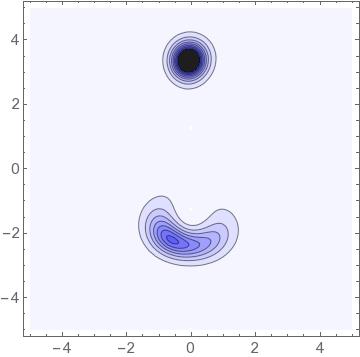}
	\caption{Plot of the topological charge density with one instanton at the origin and the other two at decreasing values of $\tau_{i}$. Note the apparent right angled scattering}
	\label{3Sca}
\end{figure}

The next step is to  analyse the dynamics numerically, as was done in the case of two instantons. Unfortunately we were unable to generate enough simulations to carry out a full analysis, however were able to observe some particular behaviours by plotting the scalar field profiles. When the instantons are far separated, this gives three peaks at the positions of each instanton, with the position defined as $\tau_{i}$ (figure \ref{Sc3Sep}). This confirms the interpretation of that parameter. If we move one instanton far away from the others (off to the right of the plot, in fact) then we see two peaks which look very similar to the graphs for two commutative instantons found in \cite{Allen2012}. The splitting in the right peak increases the closer the third instanton gets. In the graph in question, the two instantons shown are at $(\pm1,0)$ and the third is at $(0,40)$. Finally, we were able to approximate some aspects of the scattering by plotting the Topological charge Density (figure \ref{3Sca}). Here, if one instanton is kept, `stationary' at the origin, and the other instantons are plotted at successively closer values of $\tau_{i}$, there appears to be the kind of right angled scattering that is a familiar part of Soliton Dynamics. The fact that the instantons are moving away at very small values of $\tau_{i}$ is a function of the fact that the position depends both upon $\tau_{i}$ and $\sigma_{i}$, as in the two instanton case. 

\section{Conclusion}
 We have presented a new notation and method for working with noncommutative ADHM instantons, by writing them explicitly in terms of biquaternion components. After deriving the form of the ADHM equation, as well as the Moduli Space and Potential, for general SU(2) instanton number in this notation, we attempted to solve them for the two-instanton case.  \\
 First, we rederived the commutative solution which was found in \cite{Allen2012}, but using biquaternions rather than quaternions. We wereunable to find a solution on the full subspace, however we were able to find one for the subspace of the moduli space spanned by the $\mathbb{C}\times\mathbb{C}$ subalgebra of $\mathbb{C}\times\mathbb{H}$. We used this solution to calculate the metric and potential for that subspace, and checked its behaviours in various limits.  Once we had these two solutions, we investigated the dynamics on the noncommutative moduli space numerically. In general, increasing the value of the noncommutative parameter $\zeta$ increased the repulsion between the instantons, however the addition of the potential suppressed the repulsive force, particularly for large $\zeta$.\\
Finally, we looked at the case of three $U(2)$ instantons. Again, we were able to find a solution on the submanifold of the moduli space spanned by the $\mathbb{C}$ subgroup of $\mathbb{H}$. This solution once more allowed us to calculate the metric and potential on that submanifold, and to numerically graph the scalar field profiles. These solutions were very challenging numerically -- however the limited results we got indicated the presence of right angled scattering, and the appropriate behaviour of the solution in various limits. 
In terms of further work, the most obvious thing to do is to try and improve the efficiency of the numerical evaluations so that we can explore the dyonic six parameter case for  the noncommutative two instantons, and to access more of the three particle scattering in the three instanton case. Analytically, we could try and extend our ADHM solutions from the subspaces of the moduli spaces to the full moduli spaces. 

This would allow us to use the moduli space to calculate the number of BPS states in this topological sector, by calculating a specific Witten index \cite{Witten1982} on the moduli space. As argued in \cite{Witten1982} the index is invariant under changes of the parameters of a theory, provided that these changes can be expressed as a conjugation of the supersymmetric variables in the theory, as such a conjugation leaves the supersymmetry properties of a theory unchanged.  The moduli space potential is one of these parameters. \\
In \cite{Stern2000} it is argued that the Dirac operator on the moduli space goes as $\text{exp}(-q)$, where $q$ is the absolute value of the moduli space potential, outside of small regions around the zeros of that potential. It is shown that rescaling the potential does not change the Witten index, so the number of BPS states remains the same in the limit that $q\rightarrow \infty$. In this limit the moduli space dynamics are exponentially suppressed outside of the small regions around the zeros. On the commutative moduli space this procedure cannot be carried out due to the presence of singularities, but because these singularities are removed by the multiplicity, in this case these regions should be describable by supersymmetric harmonic oscillators, as was found to be the case in \cite{Bak2013} for single U(2) instantons. Calculating the Witten index, and hence the number of BPS states, would then be fairly straightforward. Therefore a key goal for future research would be to find the full moduli space and take the potential to infinity. This would suppress the dynamics everywhere except from regions of a small radius around the zeros of the potential, where it would be described by a supersymmetric quantum mechanics, for which we could calculate the Witten Index, and hence the partition function. Once we have calculated this function, we can compare it to the result directly calculated from the field theory in \cite{Kim2011}. This would provide a check for the $k=2$ case of the hypothesis mentioned in the introduction, where the (2,0) theory describing the interaction of multiple M5 Branes is the strong coupling completion of  5d Super Yang Mills. If this is true, then the instantons of topological charge $ k$  should match with states of Kaluza Klein momentum $k $ around the circle of compactification, as explained in \cite{Lambert2013}. The $k=2$ section of the partition function for Kaluza Klein states was calculated in \cite{Kim2011}, and so comparing this to the partition function on the instanton moduli space would be an important next step in verifying this conjecture. 

\appendix
\section{Solving the ADHM equations}
\label{sec:the-noncommutative-case}
As stated in equation \ref{CADHM}, the noncommutative ADHM equations are: 
\begin{align} \label{NCADHM}\nonumber
2\imh(\bar{\sigma}_{R}\sigma_{I})-\imh(\bar{w}_{R}w_{I})\imh(\bar{v}_{R}v_{I})=0 \\ \nonumber
\imh(\bar{w}_{R}w_{I})+\imh(\bar{v}_{R}v_{I})=-4\zeta\sigma_{3}\\ \nonumber
\imh(\bar{\tau}\sigma_{I})=\frac{ \imh(\bar{w}_{R}v_{I}+\bar{v}_{R}w_{I})}{2} \equiv \frac{\Upsilon}{2}\\  
\imh(\bar{\tau}\sigma_{R})=\frac{\imh(\bar{w}_{R}v_{R}+\bar{w}_{I}v_{I})}{2}\equiv\frac{\Lambda}{2} 
\end{align}
where $\sigma_{3}$ is the quaternion basis element
\begin{equation}
\begin{bmatrix}
i & 0 \\
0 & -i
\end{bmatrix}
\end{equation}
We can solve the third and fourth equations as 
\begin{equation}
\sigma_{R}=\frac{\tau}{\lvert \tau \rvert^{2}}\Big( \alpha +\frac{\Lambda}{2}\Big)
\end{equation}
and
\begin{equation}
\sigma_{I}=\frac{\tau}{\lvert \tau \rvert^{2}}\Big( \gamma +\frac{\Upsilon}{2}\Big)
\end{equation}
We can use the second equation to deduce that 
\begin{equation}\label{winc}
\bar{w}_{R}w_{I}=\beta -\bar{v}_{R}v_{I}-4i\zeta\sigma_{3}
\end{equation}
and so
\begin{equation}
w_{I}=\frac{w_{R}\beta-w_{R}\bar{v}_{R}v_{I}-4i\zeta w_{R}\sigma_{3}}{\lvert w_{R}\rvert ^{2}}
\end{equation}
We want 12 independent real parameters, or three independent quaternion ones. We therefore aimed to solve for the other parameters in terms of $v_{R}, w_{R}$ and $\tau$. We were unable to find a solution to these equations for the full biquaterion moduli space.  However were able to find a solution on a complex valued geodesic submanifold. This subspace comes from restricting the quaternions to the subspace consisting of elements $z\in\mathbb{C}$ written as $x+y\sigma_{3}$, for $x,y\in\mathbb{R}$ and $\sigma_{3}$ is given by
\begin{equation}
\sigma_{3}=\begin{bmatrix}
i&0\\0&-i
\end{bmatrix}
\end{equation}
Note that $\sigma_{3}^{2}=-\mathbb{1}_{\mathbb{H}}$, and therefore $\sigma_{3}$ can play the role of the imaginary unit. We start with the second ADHM equation, now for complex variables
\begin{equation}
\imc(\bar{v}_{R}v_{I})+\imc(\bar{w}_{R}w_{I})=-4\zeta\sigma_{3}
\end{equation}
Recall that we are using the notation $\imh$ to mean the imaginary quaternion part of an element of $\mathbb{H}$; e.g. for $q=q_{0}+\vect{q}\in\mathbb{H}$,
\begin{equation}
\imh(q)=\vect{q}
\end{equation}
On the other hand, $\imc$ takes the imaginary component of an element of $\mathbb{C}$. If $z\in\mathbb{C};\ z=a +ib$
\begin{equation}
\imc(z)=b
\end{equation}
With these definitions in mind,  we can solve this for $w_{I}$ and $v_{I}$ in terms of the other variables by finding a particular solution, then by adding the null space, found by solving
\begin{equation}
\imc(\bar{v}_{R}v_{I})+\imc(\bar{w}_{R}w_{I})=0
\end{equation}
A particular solution is given by
\begin{equation}
v_{Ip}=\frac{-2\zeta v_{R}\sigma_{3}}{\lvert v_{R}\rvert^{2}}; \ w_{Ip}=\frac{-2\zeta w_{R}\sigma_{3}}{\lvert w_{R}\rvert^{2}}
\end{equation}
We already know the solution to the null equation; it is
\begin{equation}
\tilde{v}_{I}=\frac{v_{R}}{\lvert v_{R}\rvert^{2}}\Big(\beta  -\bar{w}_{R}\tilde{w}_{I}\Big)
\end{equation}
For arbitrary real $\beta$ and arbitrary quaternion $\tilde{w}_{I}$. Therefore we have the general solution
\begin{align}\label{viwigen}
v_{I}=\frac{-2\zeta v_{R}\sigma_{3}}{\lvert v_{R}\rvert^{2}}+\frac{v_{R}}{\lvert v_{R}\rvert^{2}}\Big(\beta  -\bar{w}_{R}\tilde{w}_{I}\Big) \\ \nonumber
w_{I}=\frac{-2\zeta w_{R}\sigma_{3}}{\lvert w_{R}\rvert^{2}}+\tilde{w}_{I}
\end{align}
To complete this general solution we need to solve for $\tilde{w}_{I}$. This is done by solving the first ADHM equation
\begin{equation}
\imc(\bar{\sigma}_{R}\sigma_{I})=\imc(\bar{v}_{R}v_{I})-\imc(\bar{w}_{R}w_{I})
\end{equation}
We can use two of the symmetries in equation (\ref{eqn:symms2}) to set $\text{Re}(\bar{\tau}\sigma)=0$, by analogy to \cite{Allen2012}. This corresponds to removing any component proportional to $\tau$ from $\sigma$. If we do this, then the equation becomes 
\begin{equation}
-\frac{\imc(\Lambda\Upsilon)}{\lvert\tau\rvert^{2}}=\imc(\bar{v}_{R}v_{I})-\imc(\bar{w}_{R}w_{I})
\end{equation}
If we now restrict to the complex plane spanned by $\mathbb{1}$ and $\sigma_{3}$ the LHS becomes zero, since $\Lambda$ and $\Upsilon$ are both proportional to $\sigma_{3}$, and hence their product is real and so $\imc(\Lambda\Upsilon)=0$. Putting the solutions in (\ref{viwigen}) into the RHS we get
\begin{equation}
\imc(\bar{w}_{R}\tilde{w}_{I})=0
\end{equation}
This leads to the solution
\begin{align}\label{viwigensub}
v_{I}=\frac{-2\zeta v_{R}\sigma_{3}}{\lvert v_{R}\rvert^{2}}+Bv_{R} \\ \nonumber
w_{I}=\frac{-2\zeta w_{R}\sigma_{3}}{\lvert w_{R}\rvert^{2}}+Aw_{R}
\end{align}
for $A, B \in \mathbb{R}$. We can then use the remaining two symmetries to set $A$ and $B$ above to zero -- see the discussion below equation (\ref{eqn:symms2}). Then we get the full solution for the complex subspace
\begin{align}\nonumber
&v_{I}=\frac{-2\zeta v_{R}\sigma_{3}}{\lvert v_{R}\rvert^{2}}\\ \nonumber
&w_{I}=\frac{-2\zeta w_{R}\sigma_{3}}{\lvert w_{R}\rvert^{2}}\\ \nonumber
&\sigma_{R}=\frac{\tau\text{Im}(\bar{w}_{R}v_{R}+\bar{w}_{I}v_{I})}{2\lvert\tau\rvert^2}=\frac{(\lvert v_{R}\rvert^{2}\lvert w_{R}\rvert^{2}+4\zeta^{2})}{2\lvert\tau\rvert^{2}\lvert v_{R}\rvert^{2}\lvert w_{R}\rvert^{2}}\tau\imc(\bar{w}_{R}v_{R})\sigma_{3}\\
&\sigma_{I}=\frac{\tau\imc(\bar{w}_{R}v_{I}+\bar{v}_{R}w_{I})}{2\lvert\tau\rvert^{2}}=-\frac{\zeta(\lvert w_{R}\rvert^{2}+\lvert v_{R}\rvert^{2})}{\lvert\tau\rvert^{2}\lvert v_{R}\rvert^{2}\lvert w_{R}\rvert^{2}}\tau\imc(\bar{w}_{R}v_{R}\sigma_{3})\sigma_{3}
\end{align}

\section{Solving the Scalar Field}\label{ScalarField}
The method outlined here is mainly based on Appendix 1 in \cite{Allen2012}, generalised to the case of arbitrary non commutative instantons. That method is in turn based on \cite{Dorey1996}. It begins with the ansatz
\begin{equation}\label{SFAnsatz}
\phi=iU^{\dagger}\mathcal{A}U; \ \mathcal{A}=\begin{bmatrix} q & 0 \\ 0 & P \end{bmatrix}
\end{equation}
where $\phi$ is the scalar field we are trying to calculate and $U$ is an element of the null space of the ADHM Matrix $\Delta$. Further, $q\in u(N)$, where $N$ is the degree of the instaton gauge group, and $P\in u(k)$, where $k$ is the instanton number. In fact, $iq$ is the vev of the scalar field. For the real ADHM construction, we can use $o(k)$ rather than $u(k)$. In the biquaternion case in theory there is an additional $u(1)$, promoting the symmetry group to $u(2)$. In what follows we use our freedom to choose the vev so that it lies in the $su(2)$ part of this overall $u(2)$.  \\
Note that the equation for $\phi$ has the form of a rotation of $\mathcal{A}$ by $U$. We can think of this as follows. The matrix $\mathcal{A}$ belongs in $u(N)\times u(k)$. We can imagine it being defined on a $u(N)\times u(k)$ bundle over $\mathbb{R}^{4}$.  However we know the ADHM construction breaks the, `gauge group' $u(N)\times u(k)$ down to $u(N)$. We can therefore see the rotation as rotating $\mathcal{A}$ into the $u(N)$ subspace picked out by the ADHM constraints. This interpretation can be confirmed by the straightforward observation that $U^{\dagger}(\mathbb{1}-UU^{\dagger})\mathcal{A}U=0$. A long and algebraic justification for the ansatz is given in \cite{Dorey1996}. 
Regardless of the justification for the ansatz, once we have it, the problem of solving for $\phi$ becomes the problem of solving for $P$ above. It is shown in \cite{Dorey1996} that the equation of motion for $\phi$
\begin{equation}
D^{2}\phi=0
\end{equation}
expands as
\begin{equation}\label{scalar1}
D^{2}\phi=-4iU^{\dagger}\{bfb^{\dagger}, \mathcal{A}\}U+4iU^{\dagger}bf\Trr(\Delta^{\dagger}\mathcal{A}\Delta)fb^{\dagger}U=0
\end{equation}
Here, $\Trr$ refers to the quaternion trace on each element of a matrix, not to the trace of the matrix itself, which is written $\Tr$. Hence, applied to a (complex/ real) quaternion valued matrix, $\Trr$ will give a complex/ real valued matrix, whereas $\Tr$ will give a (complex/real) quaternion.\\
With $\mathcal{A}$ written as above, the first term is $-4iU^{\dagger}\{f,p\}U$. For the second term, we recall that $\Delta$ can be written as
\begin{equation}
\Delta=\begin{bmatrix}
\Lambda\\ \Omega-\mathbb{1}x\end{bmatrix}
\end{equation}
Writing $\Omega'=\Omega-\mathbb{1}x$, recalling that $\Omega$ and $\Omega'$ are hermitian, and using the ADHM constraint $\Delta^{\dagger}\Delta=\Lambda^{\dagger}\Lambda+\Omega'^{\dagger}\Omega'=f^{-1}$ we can see
\begin{align}\nonumber
\Trr(\Delta^{\dagger}\mathcal{A}\Delta)&=\Trr(\Lambda^{\dagger}q\Lambda)+\Trr(\Omega'^{\dagger}\mathcal{A}\Omega')\\ \nonumber
&=\Trr(\Lambda^{\dagger}q\Lambda)+\frac{1}{2}\Trr\big([\Omega'^{\dagger},P]\Omega'-\omega'^{\dagger}[\Omega',P]+\{P, \Omega'^{\dagger}\Omega'\}\big)\\ \nonumber
&=\Trr(\Lambda^{\dagger}q\Lambda)+\frac{1}{2}\big([\Omega'^{\dagger}, P]\Omega'-\Omega'^{\dagger}[\Omega', P]+\{P, f^{-1}\}-\{P, \Lambda^{\dagger}\Lambda\}\big)\\ 
&=\Trr(\Lambda^{\dagger}q\Lambda)+\frac{1}{2}\big(2\Omega'^{\dagger}P\Omega'-\{\Omega'^{\dagger}\Omega',P\}+\{P, f^{-1}\}-\{P, \Lambda^{\dagger}\Lambda\}\big)
\end{align}
Now, note that $x$ in the above expression is always the coefficient of $\mathbb{1}$. Therefore the terms involving $x$ in the above expression cancel, and we can everywhere replace $\Omega'$ by $\Omega$ (this is most easily seen from the third line above).\\
We can use these to rewrite (\ref{scalar1}) as
\begin{equation}
D^{2}\phi=-4i\bigg(U^{\dagger}\{f, P-\frac{1}{2}\Trr(P)\}U+U^{\dagger}bf\Big(\Trr(\Lambda^{\dagger}q\Lambda)+\frac{1}{2}\big(2\Omega'^{\dagger}P\Omega'-\{\Omega'^{\dagger}\Omega',P\}-\{P, \Lambda^{\dagger}\Lambda\}\big)\Big)\bigg)
\end{equation}
Since $P$ has complex components, not quaternion valued ones, $\Trr(P)=P$ and the first term vanishes. Hence the e.o.m. $D^{2}\phi=0$ is equivalent to 
\begin{equation}\label{scalarmain}
\Trr(\Lambda^{\dagger}q\Lambda)+\frac{1}{2}\big(2\Omega'^{\dagger}P\Omega'-\{\Omega'^{\dagger}\Omega',P\}-\{P, \Lambda^{\dagger}\Lambda\}\big)=0
\end{equation}
This gives one equation for each component of $P$, allowing us to solve for $P$ and hence, by extension, for $\phi$. 
\section{Constructing the Moduli Space Metric and Potential}\label{MethodMetric} 
This section is based on Appendix 2 in \cite{Allen2012}, which is itself based on the method of \cite{Osborn1981} for calculating the metric determinant. This technique was adapted in \cite{Peeters2001} for the moduli space metric of two instantons, which they calculated to order $\lvert \tau\rvert^{-2}$. In \cite{Allen2012} this is extended to the full metric for 2 commutative U(2) instantons. We present the argument for arbitrary gauge group and topological charge.  \\
As in section \ref{subs:ModuliSpace}, the metric on the moduli space is defined as
\begin{equation}\label{modmetdef}
g_{rs}=\int d^{4}x\ \Tr^{\star}\big(\delta_{r}A_{i}\delta_{s}A_{i}\big)
\end{equation}
where
\begin{equation}
\delta_{r}A_{i}=\partial_{r}A_{i}-D_{i}\epsilon_{r}
\end{equation}
and
\begin{equation}\label{Trstardef}
\Tr^{\star}(q)=\Trr\big(\Tr(q)\big)
\end{equation}
There is one mode for each of the $8k$  moduli space coordinates, labelled here by the indices $r$ and $s$. The index $i$ refers to spacetime coordinates. Recall these zero modes are orthogonal to gauge transformations by definition 
\begin{equation}
D_{i}(\delta_{r}A_{i})=0
\end{equation}
We can use this fact to find an explicit expression for the metric. First, we need an expression for $\partial_{r}A_{i}|_{z=z_{0}}$ in terms of the ADHM data. To do this, we recall that $A_{i}=U^{\dagger}\partial_{i}U$, and use the identity $U=PU$ with $U$ the projection operator $\mathbb{1}-\Delta f\Delta^{\dagger}$ to derive
\begin{equation}
\partial_{r}U=-\Delta f\partial_{r}\Delta^{\dagger}U+P\partial_{r}U
\end{equation} 
Using this result, the definition $A_{i}=U^{\dagger}\partial_{i}U$, and the product rule allows us to get the necessary result 
\begin{equation}
\partial_{r}A_{i}|_{z=z_{0}}=-iU^{\dagger}\partial_{r}\Delta f\bar{e}_{i}b^{\dagger}U+iU^{\dagger}be_{i}f\partial_{r}\Delta^{\dagger}U+D_{i}(iU^{\dagger}\partial_{r}U)
\end{equation}
around an arbitrary point of the moduli space $z_{0}$. The zero mode is then this expression with the gauge dependent part removed. The third term above is explicitly a gauge transformation, however we also need to ensure that there is no gauge part implicit in the first two terms.  To do this, we use the residual transformations described in equation (\ref{symms2}) to transform the ADHM data as
\begin{equation}
\Delta\rightarrow q\Delta R, \ U\rightarrow QU, \ Q(z_{0})=\mathbb{1}, \ R(z_{0})=\mathbb{1}
\end{equation}
It can be seen that this transformation leaves $A_{i}$ invariant, and that 
\begin{equation}
\partial_{r}A_{i}|_{z=z_{0}}=-iU^{\dagger}C_{r}f\bar{e}_{i}b^{\dagger}U+iU^{\dagger}be_{i}fC^{\dagger}_{r}U+D(iU^{\dagger}\partial_{r}(Q^{\dagger}U))
\end{equation} 
with 
\begin{equation}
C_{r}=\partial_{r}\Delta+\partial_{r}Q\Delta+\Delta\partial_{r}R
\end{equation}
It turns out we can choose $C_{r}$ so that the first two terms of $\delta_{r}A_{i}$ have no gauge part -- i.e. they are a zero mode. To do so we must prove the following
\begin{lemma}
	If we choose $C_{r}$ to be independent of $x$ with
	\begin{equation}\label{eq:c10}
	\Delta^{\dagger}C_{r}=(\Delta^{\dagger}C_{r})^{T\star}
	\end{equation}
	the expression
	\begin{equation}
	\partial_{r}A_{i}=-iU^{\dagger}C_{r}f\bar{e}_{i}b^{\dagger}U+iU^{\dagger}be_{i}fC^{\dagger}_{r}U
	\end{equation}
	will be a zero mode
\end{lemma}
To do this, we first note that the condition (\ref{eq:c10}) is equivalent to the two conditions
\begin{equation}
a^{\dagger}C_{r}=(a^{\dagger}C_{r})^{T\star}; \ 	b^{\dagger}C_{r}=(b^{\dagger}C_{r})^{T\star}
\end{equation}
and then consider the expression (forming part of $\delta_{r}A_{i}$ above)
\begin{equation}
a_{i}:=U^{\dagger}bfe_{i}
\end{equation}
We can then calculate
\begin{equation}\label{aj1}
D_{i}a_{j}=\partial_{i}a_{j}-iA_{i}a_{j} \ =U^{\dagger}e_{i}bf\Delta^{\dagger}bfe_{j}+U^{\dagger}bf(\bar{e}_{i}b^{\dagger}\Delta+\Delta^{\dagger}be_{i})
\end{equation}
We then write $\Delta^{\dagger}b$ in terms of its quaternion components as $c_{k}\bar{e}_{k}$, where the $c_{k}$ are complex valued matrices. It is important to note that since $\Delta^{\dagger}b=\Omega$, the bottom $2k\times 2k$ part of the ADHM data, the $c_{k}$ are hermitian, since $\Omega$ is hermitian by construction.  Keeping this fact in mind, we can write (\ref{aj1}) as
\begin{equation}
D_{i}a_{j}=U^{\dagger}bfc_{k}f(e_{i}\bar{e}_{k}e_{j}+\bar{e}_{i}e_{k}e_{j}+\bar{e}_{k}e_{i}e_{j})
\end{equation} 
Now, we use the identity $\bar{e}_{i}e_{j}=-\bar{e}_{j}e_{i}+2\delta_{ij}$ to get
\begin{equation}
D_{i}a_{j}=-U^{\dagger}bfc_{k}(e_{i}\bar{e}_{j}e_{k}-2\delta_{jk}e_{i}-2\delta_{ik}e_{j})
\end{equation}
Then we can see $a_{j}$ satisfies both the linear selfdual field equation 
\begin{equation}
D_{[i}a_{j]}=\frac{1}{2}\epsilon_{ijkl}a_{k}a_{l}
\end{equation}
and the zero mode condition $D_{i}a_{i}=0$.
What does this say about the full mode $\delta_{r}A_{i}$? We calculate
\begin{align}\label{DdA}\nonumber
D_{i}(\delta_{r}A_{j})&=-iD_{i}U^{\dagger}C_{r}D_{i}a_{j}^{\dagger}+ia_{j}C_{r}^{\dagger}D_{i}U-iU^{\dagger}C_{r}(D_{i}a_{j})^{\dagger}+iD_{i}a_{j}C_{r}^{\dagger}U\\ \nonumber
&=-iU^{\dagger}bf\big(e_{i}\Delta^{\dagger}C_{r}\bar{e}_{j}-e_{j}C_{r}^{\dagger}\Delta\bar{e}_{i}\big)fb^{\dagger}U-iU^{\dagger}C_{r}D_{i}a_{j}^{\dagger}+iD_{i}a_{j}C_{r}^{\dagger}U\\ &-iU^{\dagger}C_{r}D_{i}a_{j}^{\dagger}+iD_{i}a_{j}C_{r}^{\dagger}U
\end{align} 
Here we have used the fact that
\begin{equation}
D_{i}U^{\dagger}-iA_{i}U^{\dagger}=U^{\dagger}e_{i}bf\Delta^{\dagger}
\end{equation}
The discussion above of $D_{i}a_{j}$ shows that the last two terms of (\ref{DdA}) are a zero mode. We must therefore check the first two terms. The only parts of these which depend on the moduli space coordinates are
\begin{equation}
e_{i}\Delta^{\dagger}C_{r}\bar{e}_{j}-e_{j}C_{r}^{\dagger}\Delta\bar{e}_{i}\equiv K_{ij}
\end{equation}
So the first two terms being a zero mode are equivalent to 
\begin{equation}
K_{[ij]}=\frac{1}{2}\epsilon_{ijkl}K_{kl}\ ;  \ K_{ii}=0
\end{equation}
and these are satisfied iff $\Delta^{\dagger}C_{r}=(\Delta^{\dagger}C_{r})^{T\star}$. This proves the above lemma. To use this result, we must see what this condition says about the form of $C_{r}$.
First we define
\begin{align}
C_{r}&=\partial_{r}\Delta+\partial_{r}Q\Delta+\Delta\partial_{r}R\\ 
Q&=\begin{bmatrix}
q & 0\\ 0 & R^{-1} \end{bmatrix}
\end{align}
Note that we can write $C_{r}$ as
\begin{equation}
\partial_{r}a+\partial_{r}Qa+a\partial_{r}R+\Big(\partial_{r}b+\partial_{r}Qb+b\partial_{r}R\Big)x
\end{equation}
Next we set $q=1$. This means it does not contribute to the variation of $Q$, which means
\begin{equation}
\partial_{r}Q=-b\partial_{r}Rb^{\dagger}
\end{equation}
(The conjugation by $b$ is necessary to give $\partial_{r}Q$ the correct dimensions). Then we see that the part of $C_{r}$ proportional to $x$ is zero, since $\partial_{r}b$ is zero as $b$ is a constant matrix and the other two terms cancel. This leaves us with 
\begin{equation}
C_{r}=\partial_{r}a+\partial_{r}Qa+a\partial_{r}R
\end{equation} 
With this form, and the fact that $R^{T\star}=-R$, since $R$ is antiunitary, it is straightforward that $b^{\dagger}C_{r}=(b^{\dagger}C_{r})^{T\star}$. The second condition, $a^{\dagger}C_{r}=(a^{\dagger}C_{r})^{T\star}$, is satisfied iff
\begin{equation}\label{Crdef}
a^{\dagger}\partial_{r}a-(a^{\dagger}\partial_{r}a)^{T\star}-a^{\dagger}b\partial_{r}Rb^{\dagger}a-(a^{\dagger}b\partial_{r}Rb^{\dagger}a)^{T\star}+a^{\dagger}a\partial_{r}R-(a^{\dagger}b\partial_{r}Rb^{\dagger}a)^{T\star}=0
\end{equation}
We have therefore reduced the problem of finding the zero modes to solving the above equation. \\
The metric is then derived from the inner product of two zero modes. To find this, we use the following result from (\cite{Osborn1981})
\begin{equation}
\Tr^{\star}(\delta_{r}A_{i}\delta_{s}A_{i})=-\frac{1}{2}\partial^{2}\Tr^{\star}\big(C_{r}^{\dagger}PC_{s}f+fC_{r}^{\dagger}C_{s}\big)
\end{equation}
where $P=\mathbb{1}-\Delta f\Delta^{\dagger}$. We can then use Stoke's Theorem to find the metric
\begin{align}\nonumber
g_{rs}&=-\frac{1}{2}\int_{\mathcal{M}} \ \partial^{2}\Tr^{\star}\big(C_{r}^{\dagger}PC_{s}f+fC_{r}^{\dagger}C_{s}\big)\\ \nonumber
&=\int_{\partial\mathcal{M}}\ \Tr^{\star}\big(C^{\dagger}_{r}P_{\infty}C_{s}+C^{\dagger}_{r}C_{s}\big)_{ij}\\ \nonumber
&=2\pi^{2}\Tr^{\star}\big(C^{\dagger}_{r}P_{\infty}C_{s}+C^{\dagger}_{r}C_{s}\big)_{ij}\\ 
&=2\pi^{2}\Tr^{\star}\Big(\partial_{r}a^{\dagger}(1+P_{\infty})\partial_{s}a-\big(a^{\dagger}\partial_{r}a-(a^{\dagger}\partial_{r}a)^{T}\big)_{ij}\partial_{s}R\Big)
\end{align}
Here
\begin{align}\nonumber
P_{\infty}&=lim_{x\rightarrow\infty}P=\mathbb{1}_{n+2k\times n+2k}-bb^{\dagger}\\ 
=&\begin{bmatrix}
\mathbb{1}_{n/2\times n/2}&0\\
0& 0_{k\times k}
\end{bmatrix}
\end{align}
remembering
\begin{equation}
\Delta(x)=\begin{bmatrix}
\Lambda\\
\Omega+\tilde{\rho}\mathbb{1}_{k\times k}
\end{bmatrix}-x
\begin{bmatrix}
0\\
\mathbb{1}_{k\times k}
\end{bmatrix}
\end{equation}
(Note -- the term in $\tilde{\rho}$ gives the center of mass, and is usually absorbed into the $x$ component by a suitable choice of coordinates, but it is there, and therefore we consider it here -- albeit briefly). The first term above then gives
\begin{equation}
2\pi^{2}\Tr^{\star}\big(da^{\dagger}(1+P_{\infty})da\big)=2\pi^{2}\Tr\big(2\Lambda^{\dagger}\Lambda+\Omega^{\dagger}\Omega+2d\tilde{\rho}^{\dagger}d\tilde{\rho}\big)
\end{equation}
The $d\tilde{\rho}^{\dagger}d\tilde{\rho}$ directions are flat and decouple from the rest of the metic and so we ignore them (They correspond to the position of the centre of mass). This gives the first part of the metric
\begin{equation}
ds_{1}^{2}=2\pi^{2}\Tr^{\star}\big(da^{\dagger}(1+P_{\infty})da\big)=2\pi^{2}\Tr^{\star}\big(2\Lambda^{\dagger}\Lambda+\Omega^{\dagger}\Omega\big)
\end{equation}
Now for the second part of the metric
\begin{equation}\label{eqn:met1}
ds_{2}^{2}=2\pi^{2}\Tr^{\star}\Big(\big(a^{\dagger}da-(a^{\dagger}da)^{T\star}\big)dR\Big)
\end{equation}
To find an explicit expression here, we write $dR$ in terms of its components considered as a $U(k)$ matrix, and solve for them using (\ref{Crdef}). We get one equation for each component, and solving them gives $dR$ in terms of the ADHM parameters. We will see this explicitly in the specific cases below. Once we have done this, we can put all these parts together to get the full metric
\begin{align}\label{eqn:met2}\nonumber
ds^{2}=ds_{1}^{2}+\ ds_{2}^{2}&=2\pi^{2}\Big(\Tr^{\star}\big(da^{\dagger}(1+P_{\infty})da\big)\\  &=2\pi^{2}\bigg(\Tr^{\star}\big(2d\Lambda^{\dagger}d\Lambda+d\Omega^{\dagger}d\Omega\big)+\Tr^{\star}\Big(\big(a^{\dagger}da-(a^{\dagger}da)^{T\star}\big)dR\Big)\bigg)
\end{align}

\section{Constructing the Potential}\label{MethodPotential2}
We can use the Metric to calculate the potential for the dyonic instanton moduli space. This also makes use of the solution for the scalar field in Appendix \ref{ScalarField}. Recall 
the definition of the potential
\begin{equation}
\mathcal{V}=\int d^{4}x\ \Tr(D_{i}\phi D_{i}\phi) 
\end{equation}
Integrating by parts and using the fact that $D^{2}\phi=0$ via its equation of motion we get
\begin{equation}
\mathcal{V}=\text{lim}_{R\rightarrow\infty}\int_{\lvert x\rvert=R}dS^{3}\hat{x}_{i}\Tr(\phi D_{i}\phi)
\end{equation}
Using the facts that $\phi=U^{\dagger}\mathcal{A} U, D_{i}=\partial_{i}-iA_{i}$ and $A_{i}=U^{\dagger}\partial_{i}U$, a moderately long calculation \cite{Allen2012} gives
\begin{equation}\label{dphi}
D_{i}\phi=iU^{\dagger}e_{i}bf\Delta^{\dagger}\mathcal{A}U+iU^{\dagger}\mathcal{A}\Delta f\bar{e}_{i}b^{\dagger}U
\end{equation}
To fully evaluate this integral, we need an expression for $U$. In general this would be rather complicated, however we only need the value of $U$ on the boundary, in the limit $R\rightarrow\infty$. For a general ADHM matrix
\begin{equation}
\begin{bmatrix}
v_{1}& v_{2} & v_{3} & \dots & v_{k}\\
\tau_{1}-x& \sigma^{\star}_{1}& \sigma^{\star}_{2}&\dots&\sigma^{\star}_{k-1}\\
\sigma_{1} & \tau_{2}-x &\sigma^{\star}_{k} &\dots & \sigma^{\star}_{2k-3}\\
\vdots& & \ddots & & \vdots\\
\sigma_{k-1}& \sigma_{2k-3}& \sigma_{3k-4} &\dots&\tau_{k}-x
\end{bmatrix}
\end{equation} the condition $\Delta^{\dagger}U=0$ is solved to leading order in $\lvert x\rvert$ by
\begin{equation}
U_{1}\mapsto 1; \ U_{i}\mapsto\frac{x}{\lvert x\rvert^{2}}v^{\dagger}_{i-1}, i\neq 1
\end{equation}
We might worry here about the issue discussed in section \ref{ADHM}, where $U$ may or may not satisfy the completeness relation (\ref{ADHMcomp}). In general we would need to worry about this, however if we expand in powers of $\zeta$, any terms including a correction of order $\zeta^{n}$ would, by dimensional analysis, also have to go as $\lvert x\rvert^{-2n}$, and are therefore neglected in this limit.
We also need these two results for the behaviour of other quantities in this limit
\begin{align}\nonumber
\Delta\mapsto \begin{bmatrix}
\Lambda\\
-x\mathbb{1}_{k}
\end{bmatrix} \\ 
f\mapsto \frac{1}{\lvert x\rvert^{2}}\mathbb{1}_{k}
\end{align}
We can use these to expand equation (\ref{dphi}), and then multiplying by $\hat{x}_{i}$ we get, to leading order
\begin{equation}
\hat{x}_{i}D_{i}\phi=\frac{2i}{\lvert x\rvert^{3}}\bigg(q\Lambda\Lambda^{\dagger}-\Lambda P\Lambda^{\dagger}\bigg)+\mathcal{O}\Big(\frac{1}{\lvert x\rvert^{4}}\Big)
\end{equation} Remembering that $\phi=iq$ on the boundary, we can then write, to leading order
\begin{align} \nonumber
\mathcal{V}&=\text{lim}_{R\rightarrow\infty}\int_{\lvert x\rvert=R}dS^{3}\hat{x}_{i}\Tr(\phi D_{i}\phi)\\ \nonumber
&=-2\text{lim}_{R\rightarrow\infty}\int_{\lvert x\rvert=R}dS^{3}\frac{1}{\lvert x\rvert^{3}}\bigg(q^{2}\Lambda\Lambda^{\dagger}-q\Lambda P\Lambda^{\dagger}\bigg)+\mathcal{O}\Big(\frac{1}{\lvert x\rvert^{4}}\Big)\\ 
&=-4\pi^{2}\Tr\Big(q^{2}\Lambda\Lambda^{\dagger}-q\Lambda P\Lambda^{\dagger}\Big)
\end{align}
Now we have these general expressions and methods for the ADHM solutions, moduli space metric, and potential, the following appendices will provide particular solutions for the cases discussed in the main text 
\section{Scalar Field, Metric and Potential in the Two Instanton case}
\subsection{The Scalar Field}
We begin with the Scalar Field. Following the method in Appendix \ref{ScalarField}, we have the Ansatz:
\begin{equation}
\phi=U^{\dagger}\mathcal{A}U; \ \mathcal{A}=
\begin{bmatrix}
q & 0\\
0 & P
\end{bmatrix}
\end{equation}
Here $q$ is in the odd graded part of $\cxh$; i.e. $q=iq_{0}+\vect{q}$, where $q_{0}\in\mathbb{R}$ and $\vect{q}\in \text{Im}_{\mathbb{H}}\mathbb{H}$.  The matrix $ P$ is antihermitian, and so can be parameterised by
\begin{equation}\label{2sfAnsatz}
\begin{bmatrix}
ai & ci-b \\
ci+b & di
\end{bmatrix}
\end{equation}
The equation for the scalar field is 
\begin{equation}
2\Trr(\Lambda^{\dagger}q\Lambda)+\Trr([\Omega^{\dagger},P]\Omega-\Omega^{\dagger}[\Omega, P])-\Trr(\{P, \Lambda^{\dagger}\Lambda\})=0
\end{equation}
Solving this equation is a lengthy calculation, which gives: 
\begin{align}\label{2sfsols}\nonumber
a=& -\frac{1}{\Theta}\bigg(A(3) \big(g^2 N_{AI}-f^2 N_{AR}\big)+A(2) w (4 g P-f N_{AR})+A(1) w
(4 f P-g N_{AI})\\ \nonumber
&-\big((16 P^2-N_{AR} N_{AI}\big) \Big(A(3) (2
s+w)+wA(4)\Big)\bigg)\\ \nonumber
b=& \frac{1}{2\Theta}\bigg(A(1) \left(f^2 (v+w)-2 N_{AI}
(\text{sv}+\text{sw}+\text{vw})\right)+A(2) (f g (v+w)+8 P
(\text{sv}+\text{sw}+\text{vw}))\\ \nonumber
&+(4 f P+g N_{AI})\Big(A(3) (v-w)-A(4) (v+w)\Big) \bigg)\\ \nonumber
c=&\frac{1}{2\Theta}\bigg(A(1)\Big( f g (v+w)-8P
(sv+sw+vw)\Big)+A(2) \Big(g^2 (v+w)+2 N_{AR} (sv+sw+vw)\Big)\\ \nonumber
&+(f N_{AR}+4 g
P)\Big(A(3) (v-w) +A(4) (v+w) \Big)\bigg)\\ \nonumber
d=& -\frac{1}{\Theta}\bigg(A(3) \big(f^2 N_{AR}-g^2
N_{AI}\big)+A(2) v (4 g P-f N_{AR})+A(1) v (4 f P-g N_{AI})\\ 
&+(A(3) (2 s+v)-A(4) v)
\big(16 P^2-X Y\big)\bigg)
\end{align}
where
\begin{align}\label{mathematicadefs}\nonumber
A(1)&=4q_{0}\reh(\bar{v}_{R}w_{I}-\bar{v}_{I}w_{R})-4\reh(\bar{v}_{R}\vect{q}w_{R}+\bar{v}_{I}\vect{q}w_{I})\\ \nonumber
A(2)&=4q_{0}\reh(\bar{v}_{R}w_{R}+\bar{v}_{I}w_{I})+4\reh(\bar{v}_{R}\vect{q}w_{I}-\bar{v}_{I}\vect{q}w_{R})\\ \nonumber
A(3)&=q_{0}\big(\lvert v_{R}\rvert^{2}+\lvert v_{I}\rvert^{2}-\lvert w_{R}\rvert^{2}-\lvert w_{I}\rvert^{2}\big)+2\reh(\bar{v}_{R}qv_{I}-\bar{w}_{R}qw_{I}) \\ \nonumber
A(4)&= q_{0}\big(\lvert v_{R}\rvert^{2}+\lvert v_{I}\rvert^{2}+\lvert w_{R}\rvert^{2}+\lvert w_{I}\rvert^{2}\big)+2\reh(\bar{v}_{R}v_{I}+\bar{w}_{R}qw_{I}) \\ \nonumber
f&=\reh(\bar{w}_{R}v_{R}+\bar{w}_{I}v_{I})   \\ \nonumber
g&= \reh(\bar{w}_{I}v_{R}-\bar{v}_{I}w_{R}) \\ \nonumber
x&=\lvert\sigma_{R}\rvert^{2} \\ \nonumber
y&=\lvert\sigma_{I}\rvert^{2} \\ \nonumber
P&=\reh(\bar{\sigma}_{R}\sigma_{I}) \\ \nonumber
v&=\lvert v_{R}\rvert^{2}+\lvert v_{I}\rvert^{2} \\ \nonumber
w&=\lvert w_{R}\rvert^{2}+\lvert w_{I}\rvert^{2} \\ \nonumber
N_{AR}&=\lvert v_{R}\rvert^{2}+\lvert v_{I}\rvert^{2}+\lvert w_{R}\rvert^{2}+\lvert w_{I}\rvert^{2}+4(\lvert\tau\rvert^{2}+\lvert\sigma_{R}\rvert^{2})  \\ \nonumber
N_{AI}&=\lvert v_{R}\rvert^{2}+\lvert v_{I}\rvert^{2}+\lvert w_{R}\rvert^{2}+\lvert w_{I}\rvert+4(\lvert\tau\rvert^{2}+\lvert\sigma_{I}\rvert^{2})\\
\Theta&= (v+w)
\big(f^2 N_{AR}-g^2 N_{AI}\big)+2 \big(16 P^2-N_{AR} N_{AI}\big) (s
v+\text{sw}+\text{vw})
\end{align}
In the commutative limit from \cite{Allen2012}, that is, $\zeta=0$ and the imaginary quaternion parts $q_{I}$ set to zero, this becomes
\begin{equation}
b=\frac{-2\reh(\bar{v}\vect{q}w)}{\Sigma_{+}+4(\lvert\tau\rvert^{2}+\lvert\sigma_{R}\rvert^{2})}; \ a, b, d=0
\end{equation}
which is precisely the result in that paper.\\
Another useful limit is that in which $\lvert\tau\rvert\mapsto\infty$. In this case
\begin{align}\label{ltlim} \nonumber
a=\frac{q_{0}(\lvert v_{R}\rvert^{2}+\lvert v_{I}\rvert^{2})+2\text{Re}(\bar{v}_{R}\vect{q}v_{I})}{\lvert v_{R}\rvert^{2}+\lvert v_{I}\rvert^{2}}
\\ 
d=\frac{q_{0}(\lvert w_{R}\rvert^{2}+\lvert w_{I}\rvert^{2})+2\text{Re}(\bar{w}_{R}\vect{q}w_{I})}{\lvert w_{R}\rvert^{2}+\lvert w_{I}\rvert^{2}}
\end{align}
With $b,c=0$. This corresponds to the two instantons being far separated. In this case we would expect them to look like two single U(2) instantons, and we see that we do in fact have two copies of the solution for a single instanton given in (\ref{sec:the-single-u2-instanton}). The next step is to use this to explicitly calculate the potential. 
\subsection{The Potential}\label{sec:2potential}
Recall from Appendix \ref{MethodPotential2} that the potential is given by
\begin{equation}
V=\int d^{4}x \Tr(D_{i}\phi D_{i}\phi)
\end{equation}
Integrating by parts, and using the equation of motion for $\phi$
\begin{equation}
D^{2}\phi=0
\end{equation}
we get
\begin{equation}\label{potdef}
V=\text{lim}_{R\mapsto\infty} \int_{\lvert x\rvert=R}dS^{3}\hat{x}_{i}\Tr\big(\phi D_{i}\phi\big)
\end{equation}
We know that the vector $U$, being a null vector of $\Delta$, must solve
\begin{align}\nonumber
v^{\dagger}U_{1}+(\tau^{\dagger}-x^{\dagger})U_{2}+\sigma^{\dagger}U_{3}=0\\ 
w^{\dagger}U_{1}+\sigma^{\dagger}U_{2}-(\tau^{\dagger}+x^{\dagger})U_{3}=0
\end{align}
This is solved on the boundary by
\begin{align}\label{umapsto}\nonumber
U_{1}\mapsto \ 1 \\ \nonumber
U_{2}\mapsto \ \frac{x}{\lvert x\rvert^{2}}v^{\dagger} \\ 
U_{3}\mapsto \ \frac{x}{\lvert x\rvert^{2}}w^{\dagger}
\end{align}
Solving these equations following the method in Appendix \ref{MethodPotential2} is long, however eventually we get:
\begin{align}\nonumber
\mathcal{V}=&8\pi^{2}\Bigg(
\lvert q\rvert^{2}\big(\lvert v_{R}\rvert^{2}+\lvert v_{I}\rvert^{2}+\lvert w_{R}\rvert^{2}+
\lvert w_{I}\rvert^{2}\big)\\ \nonumber
&
+4q_{0}\reh(\bar{v}_{R}\vec{q}v_{I}+\bar{w}_{R}\vec{q}w_{I})
-a\big(q_{0}(\lvert v_{R}\rvert^{2}+\lvert v_{I}\rvert^{2}+2\reh(\bar{v}_{R}\vect{q}v_{I})\big)\\ \nonumber
&-d\big(q_{0}(\lvert w_{R}\rvert^{2}+
\lvert w_{I}\rvert^{2})
+2\reh(\bar{w}_{R}\vec{q}w_{I})\big)+2b\reh(\bar{v}_{R}\vec{q}w_{R}+\bar{v}_{I}\vec{q}w_{I}) \\ 
&-2bq_{0}\reh(w_{I}\bar{v}_{R}-w_{R}\bar{v}_{I})-2cq_{0}\reh(w_{R}\bar{v}_{R}+w_{I}\bar{v}_{I})-2c\reh(\bar{v}_{R}\vec{q}w_{I}-\bar{v}_{I}\vec{q}w_{R})\Bigg)
\end{align}
We can choose the $q_{0}$ to be zero by requiring the vev to lie in $SU(2)$. This simplifies our solution to
\begin{align}\nonumber
\mathcal{V}=\ &8\pi^{2}\Bigg(
\lvert q\rvert^{2}\big(\lvert v_{R}\rvert^{2}+\lvert v_{I}\rvert^{2}+\lvert w_{R}\rvert^{2}+
\lvert w_{I}\rvert^{2}\big)
-a\big(2\text{Re}(\bar{v}_{R}\vect{q}v_{I})\big)-d\big(
2\text{Re}(\bar{w}_{R}\vec{q}w_{I})\big)\\ &
+2b\text{Re}(\bar{v}_{R}\vec{q}w_{R}+\bar{v}_{I}\vec{q}w_{I}) 
-2c\text{Re}(\bar{v}_{R}\vec{q}w_{I}-\bar{v}_{I}\vec{q}w_{R})\Bigg)
\end{align}
Where $a,b,c,d$ are given above. \\
If instead we go back to the large $\tau$ limit, using (\ref{ltlim})
\begin{align}\nonumber
&\mathcal{V}=8\pi^{2}\Bigg(\lvert q\rvert^{2}\big(\lvert v_{R}\rvert^{2}+\lvert v_{I}\rvert^{2}+\lvert w_{R}\rvert^{2}+
\lvert w_{I}\rvert^{2}\big)
+4q_{0}\text{Re}(\bar{v}_{R}\vec{q}v_{I}+\bar{w}_{R}\vec{q}w_{I})\\  &
-\frac{\big(q_{0}(\lvert v_{R}\rvert^{2}+\lvert v_{I}\rvert^{2}+2\text{Re}(\bar{v}_{R}\vect{q}v_{I})\big)^{2}}{\lvert v_{R}\rvert^{2}+\lvert v_{I}\rvert^{2}}-\frac{\big(q_{0}(\lvert w_{R}\rvert^{2}+
	\lvert w_{I}\rvert^{2})
	+2\text{Re}(\bar{w}_{R}\vec{q}w_{I})\big)^{2}}{\lvert w_{R}\rvert^{2}+\lvert w_{I}\rvert^{2}}
\Bigg)
\end{align}
In this case, the $q_{0}$ parts cancel explicitly, and we get
\begin{equation}\label{taulimpot}
\mathcal{V}=8\pi^{2}\lvert\vect{q}\rvert^{2}\Bigg(\hat{\vect{q}}\Big(\lvert v_{R}\rvert^{2}+\lvert v_{I}\rvert^{2}+\lvert w_{R}\rvert^{2}+
\lvert w_{I}\rvert^{2}\big)
-\frac{4\text{Re}^{2}(\bar{v}_{R}\hat{\vect{q}}v_{I})}{\lvert v_{R}\rvert^{2}+\lvert v_{I}\rvert^{2}}-\frac{4\text{Re}^{2}(\bar{w}_{R}\hat{\vect{q}}w_{I})}{\lvert w_{R}\rvert^{2}+\lvert w_{I}\rvert^{2}}
\Bigg)
\end{equation}
We would expect this is the potential for two copies of the single U(1) instanton, and if we compare to the result in section \ref{sec:the-single-u2-instanton} we can easily see that this is the case. 
\subsection{The Metric}\label{2MetApp}
As in Appendix \ref{MethodMetric}, we begin by calculating $a^{\dagger}\delta C_{r}$, and impose the condition 
\begin{equation}\label{cond1}
a^{\dagger}\delta C_{r}=\big( a^{\dagger}\delta C_{r} \big)^{T\star}
\end{equation}
Once again, we carefully note that $T$ involves taking the transpose considered as a $2\times 2$ matrix of biquaternions. It does not affect the quaternions themselves. The operation $\star$ takes the complex conjugate of each element, which again does not affect the quaternions but only their complex coefficients.\\
We can expand $\delta R$ in the $u(2)$ basis as
\begin{equation}\label{dRdef}
\delta R = \begin{bmatrix}
id\phi & id\psi-d\theta \\ id\psi+d\theta & id\chi
\end{bmatrix}
\end{equation}
this gives 3 simultaneous equations for the derivations in the different gauge directions. As discussed below equation (\ref{eqn:met1}), we can use these along with equation (\ref{Crdef}) to find $\delta R$ in full: 

\begin{align}\nonumber
d\phi=& \frac{1}{\Phi}\bigg(-2 B(1) (2 s+w) (4 f P-g N_{AI})+2 B(2) ((2 s+w) (4 g
P-f N_{AR})\\ \nonumber
&-(B(3) (2 s+w)+B(4) w) \left(16 P^2-N_{AR}
N_{AI}\right)-2 B(4) \left(f^2 N_{AR}-8 f g P+g^2 N_{AI}\right)\bigg)\\ \nonumber
d\theta=&\frac{1}{\Phi}\bigg(2 B(1) \left(f^2 (4 s+v+w)-N_{AI} (sv+sw+vw)\right)\\ \nonumber
&+2 B(2) ( f g (4 s+v+w)-4 P (sv+sw+vw))\\ \nonumber
&+(B(3) (4s+v+w)-B(4) (v-w)) (4 f P-g N_{AI})\bigg)\\ \nonumber
d\psi=&\frac{1}{\Phi}\bigg(-2 B(1) (f g (4 s+v+w)-4 P (sv+sw+ vw))\\ \nonumber
&-2 B(2) \left(g^2 (4 s+v+w)-2 N_{AR}
(sv+sw+vw)\right)\\ \nonumber
&+(B(3) (4
s+v+w)-B(4) (v-w)) (4 g P-f N_{AR})\bigg)\\ \nonumber
d\chi=&\frac{1}{\Phi}\bigg(-2 B(1)
(2 s+v) (4 f P-g N_{AI})+2 B(2) (2 s+v) (4 g P-f N_{AR})\\ &
-\left(16 P^2-N_{AR}N_{AI}\right) (B(3) (2 s+v)-B(4) v)+2 B(4) \left(f^2 N_{AR}-8 f g P+g^2 N_{AI}\right)\bigg)
\end{align}
where the terms are defined in (\ref{mathematicadefs}) with the addition of 
\begin{align}\nonumber
B(1)=&	\bar{v}_{R}dw_{R}+\bar{v}_{I}dw_{I}-\bar{w}_{R}dv_{R}-w_{I}dv_{I}+2(\bar{\tau}d\sigma_{R}-\bar{\sigma}_{R}d\tau)\\ \nonumber
B(2)=&\bar{v}_{R}dw_{I}-\bar{v}_{I}dw_{R}+\bar{w}_{R}dv_{I}-\bar{w}_{I}dv_{R}+2(\bar{\sigma}_{I}d\tau-\bar{\tau}d\sigma_{I})\\ \nonumber
B(3)=&\bar{v}_{R}dv_{I}-\bar{v}_{I}dv_{R}+\bar{w}_{R}dw_{I}-\bar{w}_{I}dw_{R}\\ \nonumber
B(4)=&\bar{v}_{R}dv_{I}-\bar{v}_{I}dv_{R}-\bar{w}_{R}dw_{I}+\bar{w}_{I}dw_{R}+2(\bar{\sigma}_{R}d\sigma_{I}-\bar{\sigma}_{I}d\sigma_{R})\\ 
\Phi=&4 \left((4 s+v+w) \left(f^2 X-8 f g P+g^2 Y\right)+\left(16
P^2-\text{XY}\right) (sv+sw+vw)\right)
\end{align}

We can now use this in our formula (\ref{eqn:met2})
\begin{equation}
ds^{2}=ds_{1}^{2}+ds_{2}^{2}=2\pi^{2}\bigg(\Tr^{\star}\big(2d\Lambda^{\dagger}d\Lambda+d\Omega^{\dagger}d\Omega\big)+\Tr^{\star}\Big(\big(a^{\dagger}da-(a^{\dagger}da)^{T\star}\big)dR\Big)\bigg)
\end{equation}
First we have that $a^{\dagger}da-(a^{\dagger}da)^{T\star}$ is
\scriptsize \begin{align}\nonumber
&\begin{bmatrix}
0 & \bar{v}_{R}dw_{R}+\bar{v}_{I}dw_{I}-\bar{w}_{R}dv_{R}-\bar{w}_{I}dv_{I}+2(\bar{\tau}d\sigma_{R}-\bar{\sigma}_{R}d\tau) \\
-\big( \bar{v}_{R}dw_{R}+\bar{v}_{I}dw_{I}-\bar{w}_{R}dv_{R}-\bar{w}_{I}dv_{I}+2(\bar{\tau}d\sigma_{R}-\bar{\sigma}_{R}d\tau) \big) & 0
\end{bmatrix} \\
&+i\begin{bmatrix}
2\big(\bar{v}_{R}dv_{I}-\bar{v}_{I}dv_{R}+\bar{\sigma}_{R}d\sigma_{I}-\bar{\sigma}_{I}d\sigma_{R}\big) &
\bar{v}_{R}dw_{I}-\bar{v}_{I}dw_{R}+\bar{w}_{R}dv_{I}-\bar{w}_{I}dv_{R} +2\big(\bar{\sigma}_{I}d\tau-\bar{\tau}d\sigma_{I}\big) \\
\bar{v}_{R}dw_{I}-\bar{v}_{I}dw_{R}+\bar{w}_{R}dv_{I}-\bar{w}_{I}dv_{R} +2\big(\bar{\sigma}_{I}d\tau-\bar{\tau}d\sigma_{I}\big) &
2\big(\bar{w}_{R}dw_{I}-\bar{w}_{I}dw_{R}-\bar{\sigma}_{R}d\sigma_{I}+\bar{\sigma}_{I}d\sigma_{R}\big)
\end{bmatrix}
\end{align}
\normalsize
Once we have this it is fairly straightforward to calculate the metric as
\begin{align}\label{2MetSol}\nonumber
ds^{2}=&8\pi^{2}\bigg(d^{2}v_{R}+d^{2}v_{I}+d^{2}w_{R}+d^{2}w_{I}+d^{2}\tau+d^{2}\sigma_{R}+d^{2}\sigma_{I}\\ \nonumber
&-\reh\Big(\big(\bar{v}_{R}dv_{I}-\bar{v}_{I}dv_{R}+\bar{\sigma}_{R}d\sigma_{I}-\bar{\sigma}_{I}d\sigma_{R}\big)d\phi+\big(\bar{w}_{R}dw_{I}-\bar{w}_{I}dw_{R}-\bar{\sigma}_{R}d\sigma_{I}+\bar{\sigma}_{I}d\sigma_{R}\big)d\chi\\ \nonumber
&+\big(\bar{v}_{R}dw_{R}+\bar{v}_{I}dw_{I}-\bar{w}_{R}dv_{R}-\bar{w}_{I}dv_{I}+2(\bar{\tau}d\sigma_{R}-\bar{\sigma}_{R}d\tau))\big)d\theta\\ 
&	+\big(\bar{v}_{R}dw_{I}-\bar{v}_{I}dw_{R}+\bar{w}_{R}dv_{I}-\bar{w}_{I}dv_{R} +2\big(\bar{\sigma}_{I}d\tau-\bar{\tau}d\sigma_{I}\big)\big)d\psi\Big)
\end{align}
We can check the behaviour of this solution in various limits. First of all, the commutative real limit, where the various imaginary quaternion parts $q_{I}$ and the noncommutative parameter $\zeta$ are set to zero. In this limit we have
\begin{equation}
d\phi=d\psi=d\chi=0; d\theta=\frac{\bar{v}_{R}dw_{R}-\bar{w}_{R}dv_{R}-\bar{w}_{R}dv_{R}+2(\bar{\tau}d\sigma_{R}-\bar{\sigma}_{R}d\tau)}{\lvert v_{R}\rvert^{2}+\lvert w_{R}\rvert^{2}+4(\lvert\tau\rvert^{2}+\lvert\sigma_{R}\rvert^{2})}
\end{equation}
This allows us to calculate the metric to be 
\begin{equation}
ds^{2}=8\pi^{2}\Big(d^{2}v_{R}+d^{2}w_{R}+d^{2}\tau+d^{2}\sigma_{R}-\frac{dk^{2}}{N_{A}}\Big)
\end{equation}
with 
\begin{align}
N_{A}&=\lvert v_{R}\rvert^{2}+\lvert w_{R}\rvert^{2}+4(\lvert\tau\rvert^{2}+\lvert\sigma_{R}\rvert^{2})\\ \nonumber
dk&=\bar{v}_{R}dw_{R}-\bar{w}_{R}dv_{R}-\bar{w}_{R}dv_{R}+2(\bar{\tau}d\sigma_{R}-\bar{\sigma}_{R}d\tau)
\end{align}
exactly as in \cite{Allen2012}. The second limit we can check is the limit in which $\lvert\tau\rvert\mapsto\infty$. Since this corresponds to the two instantons becoming far separated, in this limit, we would expect to get two copies of the solution for a single $U(2)$ instanton. We in fact get
\begin{equation}
d\phi=\frac{v_{R}dv_{I}-\bar{v}_{I}dv_{R}}{\lvert v_{R}\rvert^{2}+\lvert v_{I}\rvert^{2}};
\ d\chi=\frac{w_{R}dw_{I}-\bar{w}_{I}dw_{R}}{\lvert w_{R}\rvert^{2}+\lvert w_{I}\rvert^{2}}
\end{equation}
This gives the metric
\begin{align}
ds^{2}=8\pi^{2}\bigg(d^{2}v_{R}+d^{2}v_{I}+d^{2}w_{R}+d^{2}w_{I}-\frac{\big(v_{R}dv_{I}-\bar{v}_{I}dv_{R}\big)^{2}}{\lvert v_{R}\rvert^{2}+\lvert v_{I}\rvert^{2}}-\frac{\big(w_{R}dw_{I}-\bar{w}_{I}dw_{R}\big)^{2}}{\lvert w_{R}\rvert^{2}+\lvert w_{I}\rvert^{2}}\bigg)
\end{align}
This is precisely the sum of two copies of the form in section \ref{sec:the-single-u2-instanton} above, equation (\ref{1metricsol}). 
\section{Three Instanton Metric and Potential}\label{App:ThreeInst}
\subsection{The Scalar Field}
As in the 2 Instanton case the expressions derived here for the scalar field, metric and potential are in principle valid for the full quaternion parametrisation. However when we substitute in the solutions for the $\sigma_{i}$ derived in equation (\ref{eqn:3sigma}), that is only valid for that particular complex subspace. Keeping this in mind, we use the same method as before. This time the ansatz is given by
\begin{equation}
\mathcal{A}=\begin{bmatrix}
\vect{q} &0 \\ 0 & P
\end{bmatrix}
\end{equation}
where $\vect{q}\in\ su(2)$, and $P\in\ o(3)$, parametrised as
\begin{equation}
\begin{bmatrix}
0 & -a & b \\ a & 0 &-c \\ -b & c & 0
\end{bmatrix}
\end{equation}
The ADHM data $\Delta$ is given, in this case, by
\begin{equation}
\begin{bmatrix}
u & v & w\\
\tau_{1}&\sigma_{1}&\sigma_{2}\\
\sigma_{1}&\tau_{2}&\sigma_{3}\\
\sigma_{2}&\sigma_{3}&\tau_{3}
\end{bmatrix}
\end{equation}
where $\tau_{1}+\tau_{2}+\tau_{3}=0$. Now the elements are all quaternions, not biquaternions. The equation we want to solve is still
\begin{equation}\label{o3scaeqn}
2\Trr(\Lambda^{\dagger}\vect{q}\Lambda)+\Trr([\Omega^{\dagger},P]\Omega-\Omega^{\dagger}[\Omega, P])-\Trr(\{P, \Lambda^{\dagger}\Lambda\})=0
\end{equation}
Proceeding as in the previous cases, we can solve for the components of $\mathcal{A}$ as
\begin{align}\label{3sfSol}\nonumber
a=&\frac{1}{\Upsilon}\bigg(C_3 (2 \Psi_{2} M_{A2}-\Psi_{1}\Psi_{3})+C_2 (2\Psi_{1} M_{A3}-\Psi_{2}\Psi_{3})+C_1 \left(-\left(4 M_{A2}
M_{A3}-\Psi_{3}^2\right)\right)\bigg)\\ \nonumber
b=&\frac{1}{\Upsilon}\bigg(-C_3 (\Psi_{1}\Psi_{2}+2 M_{A1} z)+C_1 (2\Psi_{1} M_{A3}-\Psi_{2}\Psi_{3})-C_2 \left(4 M_{A1} M_{A3}-\Psi_{2}^2\right)\bigg)\\ 
c=&\frac{1}{\Upsilon}\bigg(-C_3 \left(4 M_{A1}M_{A2}-\Psi_{1}^2\right)-C_2 (\Psi_{1}\Psi_{2}+2 M_{A1}\Psi_{3})+C_1 (2\Psi_{2}Y-\Psi_{1}\Psi_{3})\bigg)
\end{align}
\normalsize
Where
\begin{align}\nonumber
C_{1}=&4\reh(\bar{v}\vect{q}u)\\ \nonumber
C_{2}=&4\reh(\bar{u}\vect{q}w)\\ \nonumber
C_{3}=& 4\reh(\bar{w}\vect{q}v)\\ \nonumber
M_{A1}=&\lvert u\rvert^{2}+\lvert v\rvert^{2}+3\lvert\sigma_{1}\rvert^{2}+\Sigma^{2}+\lvert\tau_{1}-\tau_{2}\rvert^{2}\\ \nonumber
M_{A2}=&\lvert w\rvert^{2}+\lvert v\rvert^{2}+3\lvert\sigma_{2}\rvert^{2}+\Sigma^{2}+\lvert\tau_{1}-\tau_{3}\rvert^{2}\\ \nonumber
M_{A3}=&\lvert w\rvert^{2}+\lvert v\rvert^{2}+3\lvert\sigma_{3}\rvert^{2}+\Sigma^{2}+\lvert\tau_{2}-\tau_{3}\rvert^{2}\\ \nonumber
\Psi_{1}=&\reh\big(3(\bar{\tau}_{1}\sigma_{3}-\bar{\sigma}_{2}\sigma_{1})-\bar{w}v\big)\\ \nonumber
\Psi_{2}=&\reh\big(3(\bar{\tau}_{2}\sigma_{2}-\bar{\sigma}_{1}\sigma_{3})-\bar{u}w\big)\\ \nonumber
\Psi_{3}=&\reh\big(3(\bar{\tau}_{3}\sigma_{1}-\bar{\sigma}_{3}\sigma_{2})-\bar{v}u\big)\\ 
\Upsilon=&2 \left(\Psi_{1}^2M_{A3}+\Psi_{1}\Psi_{2}\Psi_{3}-4M_{A1}M_{A2}M_{A3}+M_{A1}\Psi_{3}^2+\Psi_{2}^2M_{A2}\right)
\end{align}
\subsection{The Potential}
We can now use this to calculate the potential, using the formula
\begin{equation}
V=\int d^{4}x \Tr(D_{i}\phi D_{i}\phi)
\end{equation}
We can now follow the standard method. Integrating by parts, and using the equation of motion for $\phi$
\begin{equation}
D^{2}\phi=0
\end{equation}
We get
\begin{equation}\label{2PotSol}
V=\text{lim}_{R\mapsto\infty} \int_{\lvert x\rvert=R}dS^{3}\hat{x}_{i}\Tr\big(\phi D_{i}\phi\big)
\end{equation}
We know that the vector $U$, being a null vector of $\Delta$, must solve $\Delta^{\dagger} U=0$, which gives the equations
\begin{align}\nonumber
&\bar{u}U_{1}+(\bar{\tau}_{1}-\bar{x})U_{2}+\bar{\sigma}_{1}U_{3}+\bar{\sigma}_{2}U_{4}=0\\ \nonumber
&\bar{v}U_{1}+\bar{\sigma}_{1}U_{2}+(\bar{\tau}_{2}-\bar{x})U_{3}+\bar{\sigma}_{3}U_{4}=0\\ 
&\bar{w}U_{1}+\bar{\sigma}_{2}U_{2}+\bar{\sigma}_{3}U_{3}+(\bar{\tau}_{3}-\bar{x})U_{4}=0
\end{align}
These can be solved in the $\lvert x\rvert^{2}\mapsto\infty$ limit as
\begin{equation}
U_{1}\mapsto 1 \ ; \ U_{2}\mapsto\frac{x\bar{u}}{\lvert x\rvert^{2}}\ ; \ U_{3}\mapsto\frac{x\bar{v}}{\lvert x\rvert^{2}}\ ; \ U_{4}\mapsto\frac{x\bar{w}}{\lvert x\rvert^{2}}\ 
\end{equation}
We can continue to calculate the potential as in the previous cases. Using the method in Appendix \ref{MethodPotential2}  , analogously to the discussion for two instantons in Appendix \ref{sec:2potential} we get:
\begin{equation}\label{3PotSol}
\mathcal{V}=8\pi^{2}\bigg(\lvert\vect{q}\rvert^{2}\big(\lvert u\rvert^{2}+\lvert v\rvert^{2}+\lvert w\rvert^{2}\big)-2a\reh(\bar{v}\vect{q}u)-2b\reh(\bar{u}\vect{q}w)-2c\reh(\bar{w}\vect{q}v)\bigg)
\end{equation}
with $a,b,c$ given as above. 
\subsection{$O(3)$ Metric}
The final thing to calculate is the metric. As in the previous case, we need to calculate 
$a^{T}\delta C_{r}$, and impose the condition 
\begin{equation}\label{cond11}
a^{T}\delta C_{r}=\big( a^{T}\delta C_{r} \big)^{T\star}
\end{equation}
Note that here we have the operation $T$ rather than $\dagger$ as we are dealing with the usual, real quaternions rather than the complexified version. Since in this commutative 3-instanton case, the remaining symmetry is $o(3)$, we can write
\begin{equation}
\delta R=
\begin{bmatrix}
0 &-d\phi & d\theta\\
d\phi & 0 & -d\psi\\
-d\theta & d\psi & 0
\end{bmatrix}
\end{equation}
We should end up, analogously to the previous case, with three simultaneous equations.
We now follow the same method as before in Appendices \ref{MethodMetric} and \ref{2MetApp}. The solution is 
\begin{align}\label{3MetSol}
d\phi=&\frac{1}{\Xi}\bigg(D_{1} \left(M_{A2} M_{A3}+\Psi_{3}^2\right)+D_{2}
(M_{A3} \Psi_{1}-\Psi_{2}\Psi_{3})-D_{3} (M_{A2} \Psi_{2}+\Psi_{1}\Psi_{3})\bigg)\\ \nonumber
d\theta=&\frac{1}{\Xi}\bigg(-D_{1} (M_{A3}\Psi_{1}+\Psi_{2}\Psi_{3})-D_{2} \left(M_{A1}
M_{A3}-\Psi_{2}^2\right)+D_{3} (M_{A1} \Psi_{3}+\Psi_{1}\Psi_{2})\bigg)\\ \nonumber
d\psi=&\frac{1}{\Xi}\bigg(\text{D1} (\Psi_{1}\Psi_{3}-M_{A2}\Psi_{2})+D_{2} (M_{A1} \Psi_{3}-\Psi_{1}\Psi_{2})+D_{3} \left(M_{A1}M_{A2}-\Psi_{1}^2\right)\bigg)
\end{align}
where
\begin{align}\nonumber
D_{1}=&\bar{u}dv-\bar{v}du+\bar{\tau}_{1}d\sigma_{1}-\bar{\sigma}_{1}d\tau_{1}+\bar{\sigma}_{1}d\tau_{2}-\bar{\tau}_{2}d\sigma_{1}+\bar{\sigma}_{2}d\sigma_{3}-\bar{\sigma}_{3}d\sigma_{2}\\ \nonumber
D_{2}=&\bar{u}dw-\bar{w}du+\bar{\tau}_{1}d\sigma_{2}-\bar{\sigma}_{2}d\tau_{1}+\bar{\sigma}_{1}d\sigma_{3}-\bar{\sigma}_{3}d\sigma_{1}+\bar{\sigma}_{2}d\tau_{3}-\bar{\tau}_{3}d\sigma_{2}\\ \nonumber
D_{3}=&\bar{v}dw-\bar{w}dv+\bar{\sigma}_{1}d\sigma_{2}-\bar{\sigma}_{2}d\sigma_{1}+\bar{\tau}_{2}d\sigma_{3}-\bar{\sigma}_{3}d\tau_{2}+\bar{\sigma}_{3}d\tau_{3}-\bar{\tau}_{3}d\sigma_{3}\\ \nonumber
M_{A1}=&\lvert u\rvert^{2}+\lvert v\rvert^{2}+3\lvert\sigma_{1}\rvert^{2}+\Sigma^{2}+\lvert\tau_{1}-\tau_{2}\rvert^{2}\\ \nonumber
M_{A2}=&\lvert w\rvert^{2}+\lvert v\rvert^{2}+3\lvert\sigma_{2}\rvert^{2}+\Sigma^{2}+\lvert\tau_{1}-\tau_{3}\rvert^{2}\\ \nonumber
M_{A3}=&\lvert w\rvert^{2}+\lvert v\rvert^{2}+3\lvert\sigma_{3}\rvert^{2}+\Sigma^{2}+\lvert\tau_{2}-\tau_{3}\rvert^{2}\\ \nonumber
\Psi_{1}=&\reh\big(3(\bar{\tau}_{1}\sigma_{3}-\bar{\sigma}_{2}\sigma_{1})-\bar{w}v\big)\\ \nonumber
\Psi_{2}=&\reh\big(3(\bar{\tau}_{2}\sigma_{2}-\bar{\sigma}_{1}\sigma_{3})-\bar{u}w\big)\\ \nonumber
\Psi_{3}=&\reh\big(3(\bar{\tau}_{3}\sigma_{1}-\bar{\sigma}_{3}\sigma_{2})-\bar{v}u\big)\\ 
\Xi=&M_{A1}M_{A2}M_{A3}+M_{A1} \Psi_{3}^2-M_{A2} \Psi_{2}^2-M_{A3}\Psi_{1}^2
\end{align}
Once more we use our formula, modified for real quaternions 
\begin{equation}
ds^{2}=ds_{1}^{2}+ds_{2}^{2}=2\pi^{2}\bigg(\Tr^{\star}\big(2d\Lambda^{\dagger}d\Lambda+d\Omega^{\dagger}d\Omega\big)+\Tr^{\star}\Big(\big(a^{\dagger}da-(a^{\dagger}da)^{T}\big)dR\Big)\bigg)
\end{equation}
which enables us to calculate the metric as 
\begin{align}
ds^{2}=&8\pi^{2}\bigg(d^{2}u+d^{2}v+d^{2}w+d^{2}\tau_{1}+d^{2}\tau_{2}+d^{2}\tau^{3}+d^{2}\sigma_{1}+d^{2}\sigma_{2}+d^{2}\sigma_{3}
\\ \nonumber
&-\Big(D_{1}d\phi
+D_{2}d\theta+D_{3}d\psi\Big)\bigg)^{2}
\end{align}

\section{Numerical Methods}\label{App:numerics}
In this appendix we develop a new numerical algorithm for calculating particle trajectories on a curved manifold with potential, which we use to calculate numerical instanton scattering in the moduli space approximation in Section~\ref{sec:sixparamspace}
. We will attach a {\sc Mathematica} notebook containing the methods discussed in this appendix to a future arXiv submission of the paper. 
\subsection{Overview of the Numerical System}

The problem of instanton scattering in the moduli space approximation with $d$ parameters is equivalent to Newtonian mechanics on a $d$ dimensional curved manifold with metric $g_{\mu\nu}(x)$, potential $V(x)$ and Lagrangian $\cL(x,\dot{x}) = g_{\mu\nu}(x)\dot{x}^\mu \dot{x}^\nu - V(x)$. The corresponding equations of motion are \begin{equation}\label{eq:modulispaceEOM}\ddot{x}^\mu = - \Gamma^\mu_{\rho\sigma}(x)\dot{x}^\rho \dot{x}^\sigma - \partial^\mu V(x),\end{equation} where $\Gamma^\mu_{\rho\sigma} = \frac{1}{2} g^{\mu\nu}\left(\partial_{\rho}g_{\nu\sigma} + \partial_{\sigma}g_{\nu\rho} - \partial_{\nu}g_{\rho\sigma} \right)$ are the Christoffel connection coefficients on the manifold. We then solve for trajectories on the moduli space in the form $x(\tau; x_0, \dot{x}_0)$, where $\tau$ is the parameter on the worldline and $x_0,\dot{x}_0 \in \mathbb{R}^d$ are initial conditions specifying a given trajectory. Note that the calculations in the moduli space approximation used in this work return the line element $ds^2 = g_{\mu\nu}dx^\mu dx^\nu$, so we consider  $x(\tau; x_0, \dot{x}_0)$ as depending implicity on $ds^2$ and $V(x)$, and the metric $g_{\mu\nu}$ is extracted from $ds^2$ as an internal step in the calculation.


The standard approach to this problem first extracts the metric from the line element, then calculates the derivatives of metric and the potential as well as the inverse metric, and uses these to give the right hand side of equation~\eqref{eq:modulispaceEOM} explicitly in terms of the coordinates on the moduli space and their derivatives. In some simple cases the resulting differential equation can then be solved analytically, but the cases we are interested in in this work are too complex to solve analytically and so we treat~\eqref{eq:modulispaceEOM} as a numerical system. For a given set of numerical initial data the system can then be integrated numerically for example using {\tt NDSolve} in {\sc Mathematica}. This method was used to calculate commutative two instanton scattering in \cite{Allen2012}, both for a general gauge embedding where the moduli space has 6 degrees of freedom, and a simpler orthogonal gauge embedding with a 4 dimensional moduli space.

In Section~\ref{sec:twoinstatonsol} of this work we extend these results to noncommutative instatons, both with general and aribtrary gauge embeddings. 
Switching on the non-commutative parameter $\zeta$ or adding in more instatons both result in an increase in the algebraic complexity of the metrics and potentials on the instanton moduli space. As the complexity increases calculating the equations of motion analytically from the metric and potential in {\sc Mathematica} can become inaccessible to a standard desktop computer due the large amount of memory that the expressions require. For the two non-commutative instanton scattering with 4 dimensional moduli space considered in Section~\ref{sec:fourparamspace} it is possible to calculate the equations of motion explicitly, and so in this case we use the standard approach to integrate equation~\eqref{eq:modulispaceEOM} numerically.

For the scattering of two noncommutative instantons with 6 dimensional moduli space considered in Section~\ref{sec:sixparamspace} 
we find that it is no longer possible to calculate the equations of motion analytically due to the increased complexity of the functions on the moduli space. To illustrate this, we use the function {\tt LeafCount} as a simple measure of the complexity of an algebraic expression in {\sc Mathematica}. The {\tt LeafCount}s of line elements, metrics and potentials for the cases considered in \cite{Allen2012} and all cases considered in this work are shown in Figure~\ref{fig:leafcounts}; the cases which require the new algorithm have significantly higher {\tt LeafCount}s. A method to overcome this problem by saving large algebraic expressions to the hard disk is presented in~\cite{ISKAUSKAS:2015hla} and was used to calculate the scattering of two noncommutative instantons in \cite{Iskauskas2015}. As explained in Section~\ref{sec:twoinstatonsol} we have found some difficulties with the results presented in \cite{Iskauskas2015}, and we were not able to integrate equation~\eqref{eq:modulispaceEOM} using the method from~\cite{ISKAUSKAS:2015hla} for the updated moduli space metrics and potentials that we derive in this paper. This motivates us to develop a new algorithm which integrates equation~\eqref{eq:modulispaceEOM} numerically without explicitly calculating the equations of motion. 



\begin{figure}[h]
\begin{center}
 \begin{tabular}{|c c c c c c c|} 
 \hline
 Number of  & Commutative &  $d$ &  & {\tt LeafCount} &  & Algorithm Used\\ [0.5ex] 
  Instantons &   &    & $V(x)$ & $ds^2$ & $\frac{1}{d^2}\sum_{\mu,\nu}g_{\mu\nu}(x)$ & \\
 \hline
 2 & Yes & 4 & 35 & 105 & 10.0 & Standard\\ 
  \hline
 2 & Yes & 6 & N/A & &  & Standard\\ 
 2 & No & 4 & 255 & 715 & 70.6 & Standard\\
  \hline
 2 & No & 6 & ? & $6.9\times 10^5$ &$7.4\times 10^5$ & New\\
 \hline
\end{tabular}
\end{center}

    \caption{Algebraic complexity of moduli space potentials and line elements and average complexity of metric components for different instanton scattering problems measured by {\tt LeafCount} in Mathematica} 
    \label{fig:leafcounts}
\end{figure} 
\subsection{New Numerical Algorithm}

We now address these difficulties in calculating the equations of motion explicity analytically for moduli space metrics and potentials with complex algebraic expressions by developing a new numerical algorithm. Our key insight is to treat the right hand side of equation~\eqref{eq:modulispaceEOM} as a numerical function which calculates the partial derivatives on the manifold using a finite difference scheme. We then re-write equation~\eqref{eq:modulispaceEOM} approximately as \begin{equation}\label{eq:numericalEOM}\ddot{x}^\mu + f^\mu_{p,\delta x}(x,\dot{x})=0,\end{equation} with $f^\mu_{p,\delta x}$ defined as $$f^\mu_{p,\delta x}(x,\dot{x}) := g^{\mu\nu}(x)\left(\partial^{p,\delta x}_{\rho}g_{\nu\sigma}(x)\dot{x}^\rho \dot{x}^\sigma  -  \frac{1}{2}\partial^{p,\delta x}_{\nu}g_{\rho\sigma}(x)\dot{x}^\rho \dot{x}^\sigma  + \partial^{p,\delta x}_{\nu} V(x)\right),$$  where $\partial^{p,\delta x}_{\mu}$ are numerical partial derivatives which differ from $\partial_\mu$ with error $\cO(\delta x^p)$ which we define below.

The function $f^\mu_{p,\delta x}$ depends implicitly on the metric and potential, which must be evaluated many times at different points on the manifold to solve for a given trajectory $x(\tau; x_0, \dot{x}_0)$. As {\sc Mathematica} is an interpreted language, its default methods for evaluation of numerical functions are generally less efficient than compiled code. For this reason we use the {\sc Mathematica} function {\tt Compile} to compile the metric and potential down to C code, which signficiantly increases efficiency. In the attached {\sc Mathematica} code we then define a function {\tt NDSolveParticleTrajectory} which takes as arguments a compiled metric and potential, initial data, maximum integration time and specification for the numerical derivative, and returns a numerical trajectory $x(\tau; x_0, \dot{x}_0)$ which solves equation~\eqref{eq:numericalEOM}. In this algorithm equation~\eqref{eq:numericalEOM} is input directly into {\tt NDSolve} to take advantage of {\sc Mathematica}'s differential equation solving algorithm, which for example has inbuilt error checking and avoids the necessity of writing a Runge-Kutta algorithm to integrate in $\tau$. 

There are a number of different steps which go into the construction of  $f^\mu_{p,\delta x}$ to ensure efficient evaluation of the function for a specified numerical $x$ and $\dot{x}$. Firstly the metric must be extracted from the line element and the result compiled. For the line elements considered in~\cite{Allen2012} and Section~\ref{sec:fourparamspace}, it is simple to extract the metric from the line element directly and then work with the components of the metric, which was the method used in~\cite{Allen2012}. For the scattering problems from Sections~\ref{sec:sixparamspace} 
the average complexity of each individual component of the metric is higher than that of the entire line element, as illustrated in Figure~\ref{fig:leafcounts}. In these cases extracting the metric analytically from the line element results in an increase of memory used of around $\cO(d^2)$ which in turn results in significantly longer runtimes, and so we improve efficiency by calculating the line element numerically for each position on the manifold, and then extracting the metric from this expression. This step is encoded in the function {\tt LineElementToCompiledMetric}, which takes as arguments a line element and a list of the variables used as the coordinates on the manifold and returns a {\tt CompiledFunction} which takes a numerical point $x \in \mathbb{R}^d$ and returns $g_{\mu\nu}(x)$ as a rank 2 array. The potential is also compiled using the function {\tt CompilePotential}, which is used in a similar way.

The next step is to calculate the numerical derivatives of the metric and potential, which can be considered as a rank 2 and a rank 0 array, so we consider $f(x)$ as a general rank $r$-array on $\mathbb{R}^d$ to cover both cases. {\sc Mathematica} has an inbuilt function {\tt ND} in the package {\tt NumericalCalculus\`{}} which works by default on scalar functions of a single variable, so this must be extended to rank-$r$ arrays on $\mathbb{R}^d$ to be used in our algorithm. The extension from scalar functions to rank-$r$ arrays is straightforward because the numerical derivative calculates only linear combinations of the function, so the rank-$r$ array can be treated the same as a scalar function when calculating the numerical derivative. Extending from a single variable to calculating partial derivatives on $\mathbb{R}^d$ is more complex; {\tt ND} can be used to calculate numerical partial derivatives for functions of many variables by nesting the calculation over each dimension of the space, however this results in exponentially many function evaluations of $f(x)$ when calculating $\partial^{p,\delta x}_{\mu}f(x)$ which is not practical for our numerical algorithm. For this reason we wrote a new function {\tt CompiledND} for calculating numerical derivatives of {\tt CompiledFunctions}, which evaluates $f(x)$ only $d\, p + 1$ times when approximating $\partial^{p,\delta x}_{\mu}f(x)$.

{\tt CompiledND} calculates a $p^{\mathrm{th}}$ order forward finite difference scheme for the derivative of a {\tt CompiledFunction} $f$ which depends on a numerical vector in $x\in\mathbb{R}^d$ and returns a rank $r$ array. Its arguments are $f$, an order $p$ and a length scale $\delta x$, and it returns a {\tt Function} which depends on $x$ and returns a rank $r{+}1$ array where the additional dimension of the array encodes the index on the partial derivatives. The numerical derivative is defined based on finite differences $x + m \,\delta x\, \hat{e}_\mu$ away from the point $x$, where $\hat{e}_\mu$ are unit vectors in the $\mu$ direction in $\mathbb{R}^d$. The differences are linearly spaced in $m \in \{1, \,...\, , p\}$ and homogeneous in that the same $\delta x$ is used for each direction on the manifold. The coefficients for this finite difference scheme are given by $a^p_m = \frac{(-1)^{m+1}}{m}\; {}^p C_m$, so that 
\begin{equation}\begin{split}
\partial^{p,\delta x}_{\mu} f(x) &:=  \sum_{m=1}^p a^p_m \frac{f(x + m \,\delta x\, \hat{e}_\mu) - f(x)}{\delta x} \\
&= \frac{1}{\delta x} \left(\sum_{m=1}^p \frac{(-1)^{m+1}}{m} \; {}^p C_m \,f(x + m\, \delta x\, \hat{e}_\mu) - f(x)\sum_{m=1}^p \frac{(-1)^{m+1}}{m} \; {}^p C_m  \right)\\
&= \partial_\mu f(x) + \cO(\delta x^{p}) .\end{split}\end{equation} Then it can be seen that the error in difference from the analytical partial derivative is $\partial^{p,\delta x}_{\mu} f(x) - \partial_\mu f(x) \in \cO(\delta x^p)$, and that only $d\, p + 1$ function evalutions of $f$ have been used.


Combining all of these parts together, the following pseudocode describes schematically how {\tt NDSolveParticleTrajectory} is defined
\vspace{.7cm}
\begin{algorithmic}
\Function{\tt NDSolveParticleTrajectory}{\var{metric}, \hspace{0.1cm}\var{potential},\hspace{0.1cm}\var{initialconditions},\var{tmax},\hspace{0.1cm}\var{p},\hspace{0.1cm}\var{dx}}
\State Define $dmetric = {\tt CompiledND}(metric, p, dx)$
\State Define $dpotential = {\tt CompiledND}(potential, p, dx)$
\State Define a temporary function to calculate $f^\mu_{p,\delta x}$;
\vspace{.2cm}\Function{\tt f}{\var{x}, \var{xdot}}
\vspace{.2cm} \State Calculate inverse metric, $g^{\mu \nu} = {\tt Inverse}(metric(x))$
\vspace{.2cm} \State Calculate derivatives of metric, $dg_{\nu \rho \sigma} = dmetric(x)$
\vspace{.2cm} \State Calculate derivatives of potential, $dV_{\nu} = dpotential(x)$
\vspace{.2cm} \State Return $g^{\mu \nu} ( dg_{ \rho\nu \sigma} \;xdot^\rho\; xdot^\sigma  - \frac{1}{2}dg_{\nu \rho \sigma}\;xdot^\rho\; xdot^\sigma +  dV_{\nu})$
\EndFunction

\vspace{.2cm}\State Return {\tt NDSolve}$\big(\ddot{x}^\mu + {\tt f}(x,\dot{x})=0, initialconditions, \; t \in [0, tmax]\big)$ 
\EndFunction
\end{algorithmic}
\vspace{.7cm}

Note that the line element and potential are compiled outside of {\tt NDSolveParticleTrajectory} so that they do not need to be recompiled each time the algorithm is run for different sets of initial conditions. The following psuedocode illustrates how to calculate the particle trajectory on the moduli space defined by a potential $V(x)$ and a line element $ds^2$, 
\vspace{.7cm}
\begin{algorithmic}
\State Input $ds^2$ into {\tt LineElementToCompiledMetric} to produce {\textit g}, a {\tt CompiledFunction} 
\State Input $V(x)$ into {\tt CompiledPotential} to produce {\textit V}, a {\tt CompiledFunction} 
\State Assign numerical initial conditions $x0$ and $xdot0$
\State Assign maximum number of integration steps, $tmax$
\State {\tt NDSolveParticleTrajectory}($g, V, \{x0, xdot0\}, tmax, p, dx$)
\end{algorithmic}

\end{document}